# Cyber-Physical Security Vulnerabilities Identification and Classification in Smart Manufacturing: A Defense-in-Depth Driven Framework and Taxonomy

**Md Habibor Rahman[a] and Mohammed Shafae[b, 1]**

[a] Department of Mechanical Engineering, University of Massachusetts Dartmouth, North Dartmouth, MA 02747, USA
[b] Department of Systems and Industrial Engineering, The University of Arizona, Tucson, AZ 85721, USA

## Abstract

The increasing cybersecurity threats to critical manufacturing infrastructure necessitate proactive strategies for vulnerability identification, classification, and assessment. Traditional approaches, which define vulnerabilities as weaknesses in computational logic or information systems, often overlook the physical and cyber-physical dimensions critical to manufacturing systems, comprising intertwined cyber, physical, and human elements. As a result, existing solutions fall short of addressing the complex, domain-specific vulnerabilities of manufacturing environments. To bridge this gap, this work redefines vulnerabilities in the manufacturing context by introducing a novel characterization based on the duality between vulnerabilities and defenses. Vulnerabilities are conceptualized as exploitable gaps within various defense layers, enabling a structured investigation of manufacturing systems. This paper presents a manufacturing-specific cyber-physical defense-in-depth model, highlighting how security-aware personnel, post-production inspection systems, and process monitoring approaches can complement traditional cyber defenses to enhance system resilience. Leveraging this model, we systematically identify and classify vulnerabilities across the manufacturing cyberspace, human element, post-production inspection systems, production process monitoring, and organizational policies and procedures. This comprehensive classification introduces the first taxonomy of cyber-physical vulnerabilities in smart manufacturing systems, providing practitioners with a structured framework for addressing vulnerabilities at both the system and process levels. Finally, we demonstrate the application of the proposed framework through an illustrative smart manufacturing system, detailing its threat model, cyber-physical defense-in-depth strategy, vulnerability identification and mapping to the attack kill chain, empirical attack analysis, and potential mitigation strategies. This work equips manufacturers with actionable insights and a robust classification scheme to proactively address the cybersecurity challenges of modern manufacturing systems.

## 1 Introduction

The growing adoption of Industry 4.0 and its related technologies has transformed manufacturing systems into interconnected and distributed smart manufacturing systems, gaining the merits of decentralized, adaptive, and data-driven decision-making, operations, and control. However, integrating digital technologies and operational technology assets has introduced and increased the cyber security threat to the critical manufacturing industry from a rapidly expanding and diverse cyberattack surface [1,2]. In manufacturing, attack surface can be defined as all known and unknown vulnerabilities and controls within the cyber and physical assets of the system [3]. The ever-growing interconnectivity across these assets, the abundance of readily available manufacturing data throughout the product life-cycle management systems, and the compounded nature of the global supply chain are key contributors to exposing these vulnerabilities to adversaries in once "secure-by-isolation" manufacturing systems [4]. Recent industry reports identified a 50% increase in industrial control system-related vulnerabilities between 2020 and 2021, whereas the overall growth rate in the number of vulnerabilities was only 0.4% [5]. These emerging and existing vulnerabilities in industrial control systems and digital manufacturing technologies have significantly increased the likelihood, impact, and risk of cyberattacks on manufacturing systems [6]. As a result, manufacturing remained the top-attacked industry worldwide for three consecutive years, when 25.7% of all cyberattacks targeted the manufacturing industry in 2023 [7].

[1] Corresponding Author. *E-mail address:* shafae1@arizona.edu.

In manufacturing systems, threat actors can target cyber and/or physical assets by exploiting one or more of the system's cyber and/or physical vulnerabilities, producing attack consequences leading to organizational risk [8]. Cyberattacks against smart manufacturing systems can be categorized by the influenced/targeted and the victimized/affected domains, resulting in four attack groups as depicted in Figure 1: Cyber-to-Cyber (C2C), Cyber-to-Physical (C2P), Physical-to-Physical (P2P), and Physical-to-Cyber (P2C) [9,10]. Consequently, attacks on manufacturing are not confined only to cyber espionage (e.g., theft of data) as opposed to traditional Information Technology (IT) systems. Cyberattacks on manufacturing systems can also victimize/target the physical manufacturing domain, including manufacturing equipment [11–13], manufactured products [6,14–16], and/or the manufacturing ecosystem and sub-systems [17–19]. These non-conventional attacks can result in system sabotage, causing physical equipment damage, operational downtime, and compromised product quality and reliability, endangering public safety and human lives [14,15,20–22]. To defend the critical manufacturing infrastructure against such high-stakes attacks, manufacturing stakeholders must assess the manufacturing-specific vulnerability landscape to understand how threat actors can infiltrate their systems and evaluate the security threat the vulnerabilities pose to those systems [23,24]. Understanding, identifying, and assessing system vulnerabilities can offer insights into potential attack surfaces in manufacturing systems and provide recommendations to design and deploy appropriate defense measures for mitigating or eliminating critical vulnerabilities. Therefore, vulnerability identification and classification have been fundamental aspects of most cybersecurity frameworks and guidelines developed to address the cybersecurity threat to critical infrastructures [25,26].

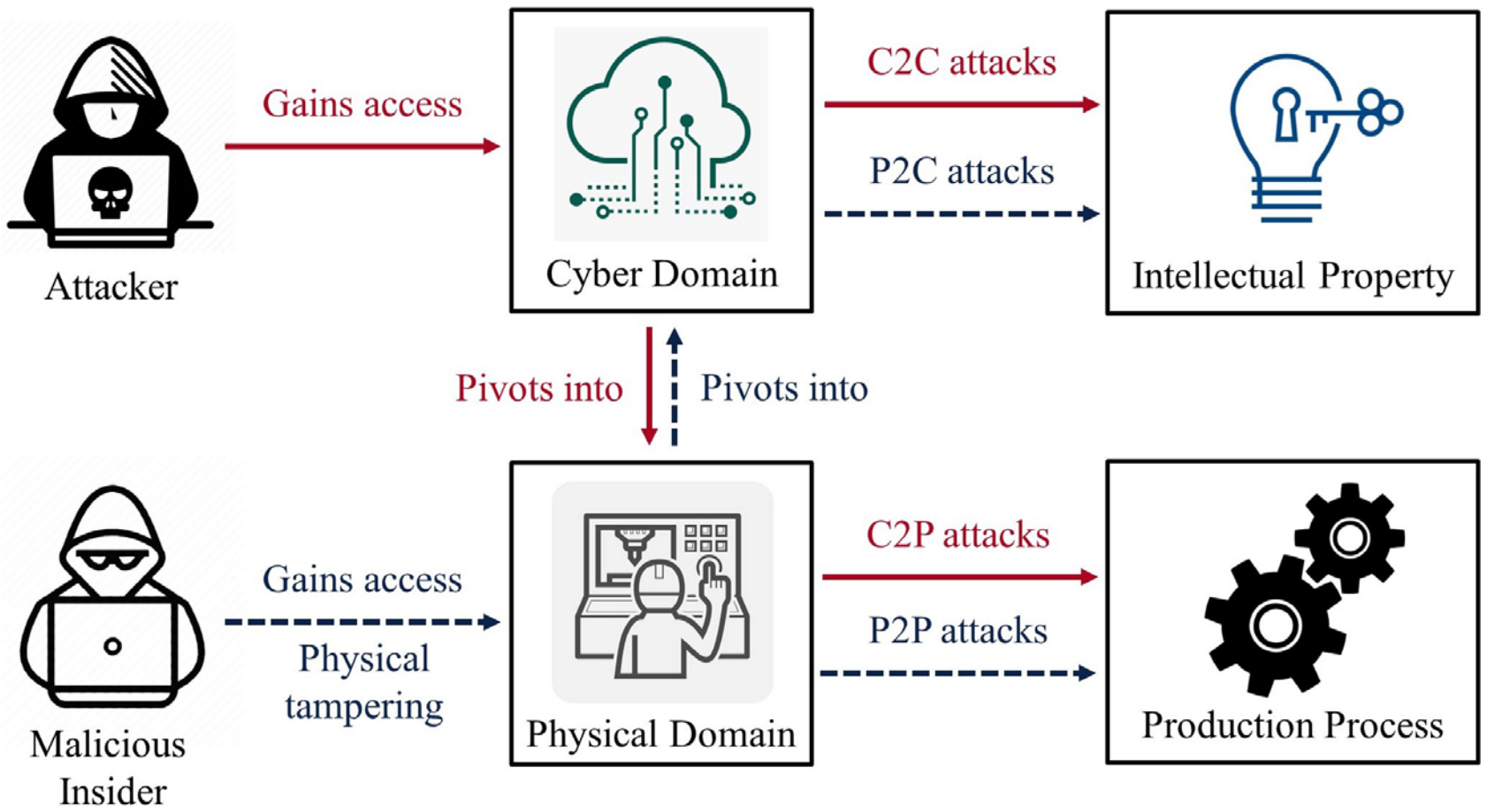

**Figure 1** Categories of cyberattacks and their impacts on smart manufacturing systems

Current related works for studying and detecting manufacturing vulnerabilities concentrate on cyber domain vulnerabilities, such as software and network vulnerabilities, and suggest mostly cyber-only protective and detective defenses to impede cyber intruders. The focus on cyber domain vulnerabilities primarily stems from (a) the traditional definition of vulnerability and (b) the guiding principle for security tools development. Security frameworks and guidelines define vulnerability as "*weakness in an information system, system security procedures, internal controls, or implementation*" [27], overlooking the physical and cyber-physical vulnerabilities in manufacturing. Additionally, traditional IT security tools and policies are defined by the CIA-triad, i.e., to ensure information Confidentiality, Integrity, and Availability [4]. Consequently, vulnerability identification and assessment are also limited to detecting and mitigating threats that can compromise data confidentiality, integrity, and availability in digital assets and network infrastructures. However, human, physical, and cyber-physical defenses can also be designed and implemented utilizing the manufacturing systems' unique characteristics and resources to augment existing cyber defenses [1,6,15,21]. For example, a successful attack (in penetrating existing cyber defenses) aiming to alter a product's geometry can still be detected at the process physical layer of the system by monitoring the manufacturing process variables, should the monitoring system have been already designed for detecting physical manifestations of cyber-physical attacks. Such monitoring capabilities or trained personnel to detect physical alterations to manufacturing parts and processes due to cyberattacks can be considered physical detective defense measures against C2P attacks on product quality [6,15]. These non-cyber defenses can also have unique vulnerabilities that threat actors can exploit to launch successful attacks on cyber-physical manufacturing systems. Currently, there is no systematic approach to understanding, identifying, and classifying potential physical and cyber-physical vulnerabilities in manufacturing systems. Therefore, we need to rethink how we define, identify, and characterize vulnerabilities in the context of manufacturing systems.

In response to the research gaps mentioned above, this work defines and characterizes cyber-physical vulnerabilities in manufacturing systems for the first time, introduces cyber-physical defense-in-depth and defense model-driven framework for vulnerability identification, and presents a comprehensive analysis of manufacturing systems' cyber, physical, and cyber-physical vulnerabilities. Additionally, taxonomical classifications are proposed to systematically characterize and categorize manufacturing-specific vulnerabilities. The specific contributions of this work are summarized as follows:

(1) This work defines manufacturing-specific cyber-physical vulnerabilities considering the unique physical and cyber-physical characteristics of smart manufacturing systems and presents the vulnerability and defense duality. Vulnerabilities are characterized by exploring what defenses are already in place, how well they are designed and deployed, and how threat actors can compromise those defense layers.

(2) It introduces the cyber-physical defense-in-depth model for manufacturing systems and explains how additional non-cyber defenses in the form of security-aware personnel, security-aware post-production inspection systems, and security-aware process monitoring approaches can complement the traditional cyber-domain defenses.

(3) Leveraging the cyber-physical defense-in-depth model and the vulnerability and defense duality, this paper identifies and classifies potential vulnerabilities in the manufacturing cyberspace, human element, post-production inspection, and production process monitoring, offering a comprehensive and structured classification scheme for manufacturing-specific vulnerabilities. This work presents the first taxonomy of cyber-physical vulnerabilities in smart manufacturing systems.

The rest of the article is organized as follows: Section 2 presents relevant works on identifying and understanding vulnerabilities and highlights their limitations. Section 3 presents the vulnerability identification approach. Specifically, Section 3.1 discusses the proposed vulnerability definition and the need to map vulnerabilities to varying stages of attacks, the concept of vulnerability and defense duality is explained in Section 3.2, and Section 3.3 introduces the cyber-physical defense-in-depth model. Leveraging the defense model and the vulnerability and defense duality, Section 3.4 presents the vulnerability identification approach. Vulnerabilities in the manufacturing cyberspace, human element, post-production inspection, and production process monitoring are categorized and compiled in Section 4, offering the first taxonomical classification of cyber-physical vulnerabilities. The proposed cyber-physical defense-in-depth model and the vulnerability characterization and identification scheme are demonstrated for an illustrative smart manufacturing system and its corresponding threat model in Sections 5.1-5.4. Section 5.5 presents several empirical cyber-physical attacks demonstrating real-world exploitation of the identified vulnerabilities, and several countermeasures for vulnerability mitigation are reported in Section 5.6. Finally, Section 6 draws the paper to its conclusion.

# 2 Related works

Current works on identifying and understanding vulnerabilities primarily fall into three categories: (1) repositories and guidelines, (2) vulnerability identification tools, and (3) academic literature on cybersecurity. This section provides a brief review of these efforts.

## 2.1 Public repositories and guidelines

There are several government-funded and community-sharing repositories for identifying and managing vulnerabilities, such as the National Vulnerability Database (NVD) [28], Common Vulnerabilities and Exposures (CVE) [29], Common Weakness Enumeration (CWE) [30], and IBM X-Force Exchange [31]. However, these repositories focus on cyber-domain vulnerabilities related to computational logic and software/IT security with little to no focus on manufacturing systems' physical and joint cyber-physical vulnerabilities. Different guidelines proposed for industrial control systems security, such as the one from the US National Institute of Standards and Technology (NIST) [32], also primarily characterize vulnerabilities in system architecture and design, software development, communication and network, and configuration and maintenance, i.e., all are vulnerabilities in the cyber domain. The discussion on physical vulnerabilities is limited to physical access control, i.e., restrictive physical access to information assets [33].

## 2.2 Vulnerability identification and assessment tools

The two key approaches for system vulnerability identification are (1) scan-based vulnerability analysis and (2) audit-based analysis [34]. Scan-based vulnerability assessment tools such as the Shodan search engine for security [35], Nmap utility for network discovery and security auditing [36], and Metasploit penetration testing software [37] are often utilized to find system vulnerabilities. These tools can automatically scan a system's network and connected devices to detect vulnerabilities [38,39] but are limited to identifying software and network vulnerabilities. On the other hand, audit-based analysis is questionnaire-driven and typically goes beyond software and network vulnerabilities to evaluate an organization's security measures and practices. The Cyber Security Evaluation Tool (CSET) is an example of such a tool created by the US Department of Homeland Security [40]. CSET questionnaire asks detailed questions about system components and architectures, operational policies, and procedures, providing a gradual approach to assess security measures and practices. Comparing the answers from the questionnaire to the requirements identified in industry-recognized standards, CSET offers a scheme of strengths and weaknesses and a catalog of recommended actions for improving the organization's cybersecurity posture. However, the existing reference standards for the comparison are unrelated to manufacturing or do not cover the joint cyber-physical vulnerabilities of manufacturing systems.

## 2.3 Academic literature

Like existing repositories and guidelines, current academic literature on cybersecurity mainly covers cyber-domain vulnerabilities, primarily focusing on the confidentiality, integrity, and availability of sensitive digital information, Intellectual Property (IP), and trade secrets [41]. Vulnerable communication systems, poor security policies, and adoption of commercially off-the-shelf products (e.g., software, hardware) with inherent cyber vulnerabilities are considered the primary causes of manufacturing vulnerabilities [4]. As a result, manufacturing vulnerabilities are broadly categorized into data, security administration, software, operating system, and network communication system vulnerabilities. Over the last decade, the growing concern of convoluted and cascaded cyberattacks against manufacturing organizations has prompted the investigation of manufacturing-specific vulnerabilities beyond traditional network and software vulnerabilities. DeSmit et al. (2017) proposed a decision tree-based approach to assess vulnerabilities in manufacturing systems occurring at the intersections of cyber, physical, cyber-physical, and human entities [1]. They recommended several metrics, such as loss of information, inconsistency, relative frequency, lack of maturity, and time until detection, for the assessment without properly characterizing the vulnerabilities. Sturm et al. (2017) categorized cyber-physical vulnerabilities in Additive Manufacturing (AM) based on the digital AM process chain. Vulnerabilities were grouped into CAD model, STL file, tool path file, physical machine, raw material, and the process monitoring and Quality Control (QC) system, which are essentially attack locations but not vulnerabilities. Elhabashy et al. (2020) identified vulnerabilities in quality inspection systems. They categorized vulnerabilities as the improper implementation of QC tools, violation of statistical assumptions related to QC tools such as control charts[2], inadequate and infrequent data collection for inspection, and inspection of only a subset of product features [21]. However, a holistic approach to identifying manufacturing-specific vulnerabilities is still missing. Additionally, a comprehensive classification and assessment of manufacturing-specific vulnerabilities encompassing vulnerabilities in the human element, inspection system, and production process is lacking in the literature.

# 3 Vulnerability identification approach

Vulnerability identification and characterization can aid manufacturers in proactively strengthening their security posture and reducing cybersecurity risk. Toward developing a systematic vulnerability identification and classification framework, this section (1) proposes a comprehensive definition of vulnerabilities in the context of manufacturing systems and highlights the importance of mapping vulnerabilities to attack frameworks, (2) discusses the concept of vulnerability and defense duality, (3) introduces the cyber-physical defense-in-depth model for manufacturing system security, and (4) presents a structured approach for vulnerability identification.

## 3.1 Redefining vulnerabilities

**Existing vulnerability definition and its limitation**. Vulnerabilities are commonly defined as flaws in a system, and the general cybersecurity literature and guidelines primarily refer to deficiencies in computational logic or information systems. For example, NIST defines cybersecurity vulnerability as "weakness in an information system, system security procedures, internal controls, or implementation that could be exploited or triggered by a threat source" [27]. These cyber domain vulnerabilities apply to manufacturing systems – identifying, classifying, and assessing those are required for smart manufacturing systems – but this is insufficient to address the security of typically complicated manufacturing systems consisting of intertwined cyber, physical, and human elements. For instance, software security upgrades can help avoid software vulnerability exploitation, and continuous updates are affordable in traditional cybersecurity. However, such core presumptions are invalid in manufacturing due to the presence of legacy systems run by outdated software with functional physical technology considered current, which are cost-prohibitive and impractical to upgrade with security patches [42]. Manufacturing systems are also characterized by high heterogeneity between and within different groups of physical devices (e.g., machines, material handling, metrology systems, and sensors) that rely on open-by-design communication protocols for communication flexibility [4]. Additionally, human error, insider threat, higher product mixes, processing uncertainty, and complex and interdependent global supply chains distinguish manufacturing from IT/software systems and other cyber-physical systems.

**Proposed vulnerability definition**. Effectively securing manufacturing systems requires an outlook on these unique characteristics and rethinking how vulnerabilities can be identified and mitigated in the manufacturing context. As an initial step toward security-aware manufacturing system design across the product's life cycle, we extend the current vulnerability definition as *any deficiency across the product's life cycle that can be maliciously utilized to steal Intellectual Property (IP), degrade overall equipment efficiency, disrupt/sabotage production system/process safe operations, and/or tamper with the product's intended quality and functional performance*.

**Mapping vulnerabilities to attack framework**. To effectively mitigate cyber-physical threats, it is essential to understand how vulnerabilities are strategically exploited in an attack. Depending on the motivation and targets of the attack, a successful attack may involve a series of activities from attack launch to execution. Aligning vulnerabilities in manufacturing systems with different activities for attack execution will provide a better understanding of how and when threat actors exploit those vulnerabilities. For example, the cyber kill chain framework developed by Lockheed Martin describes successful cyberattacks as progression through seven stages: (1) reconnaissance, where attackers identify vulnerabilities; (2) weaponization, developing malware or exploit payloads; (3) delivery, transmitting the attack through phishing, infected files, or network intrusions; (4) exploitation, triggering vulnerabilities to gain unauthorized access; (5) installation, embedding persistence mechanisms such as backdoors or rootkits; (6) command & control, establishing remote access for manipulation; and (7) actions on objectives, executing final goals such as data theft, sabotage, or tampering with the product and/or process [43]. Each stage may require exploiting different vulnerabilities as adversaries advance through the attack chain to ensure success. Therefore, mapping system vulnerabilities to varying stages of attacks can help reveal the paths that threat actors use to compromise manufacturing system assets.

[2] A control chart monitors process stability by detecting special-cause variations that indicate deviations from normal conditions. It plots time-ordered data with a center line representing the process mean and control limits based on expected process variation [174]. Observations outside these limits signal potential instability, requiring corrective action.

### 3.2 Vulnerability and defense duality

This work characterizes system vulnerabilities as *the absence of proper defense strategies and/or measures*. Cybersecurity risk in manufacturing systems arises from threat events when threat actors target specific locations in the manufacturing value chain and exploit system vulnerabilities, leading to damages [24]. Understanding and identifying vulnerabilities is critical to secure the manufacturing industry, as highlighted in a high-level proactive security model presented in Figure 2. The identified vulnerabilities can guide the risk assessment procedure, which can then offer insights into designing and implementing effective countermeasures so that vulnerabilities can be removed or mitigated [44,45]. Fewer vulnerabilities can be exposed when appropriate countermeasures (also known as defenses) are developed and deployed in a system. However, a lack of adequate defense measures or their improper implementation will result in failure to eliminate or mitigate system vulnerabilities, and therefore, vulnerabilities persist in the system and increase the attack surface. Following this notion of vulnerability and defense duality, one way to investigate the manufacturing system's vulnerabilities is to explore what defenses are already in place and how well they are designed and deployed.

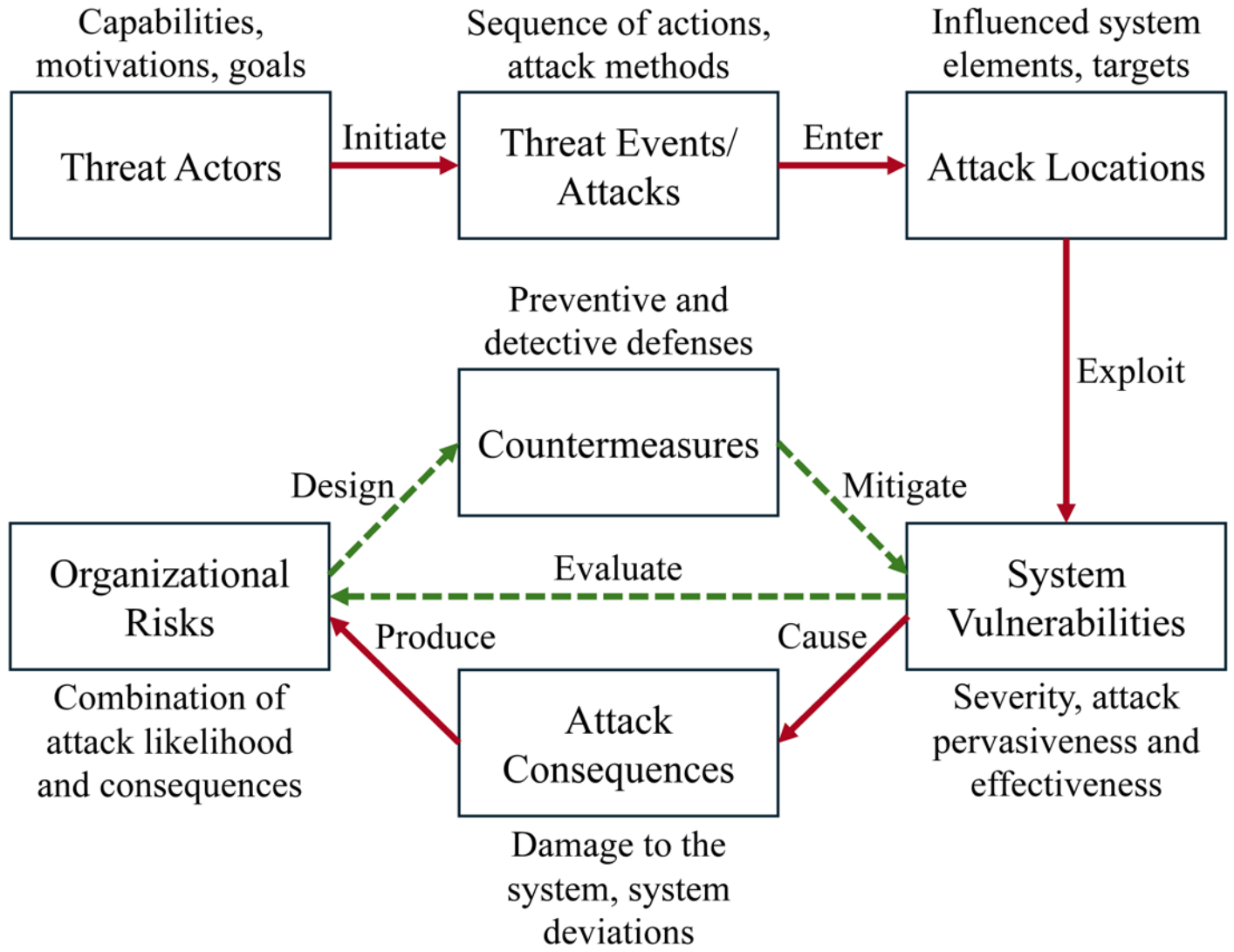

**Figure 2** Cybersecurity risk model for smart manufacturing systems and a high-level proactive security model (shown by the dashed lines)

This work first explores what constitutes a complete and thorough defense scheme for manufacturing systems to drive the investigation of system-level vulnerabilities. The defense-in-depth security model, for example, is an approach toward realizing appropriate and comprehensive defenses for a system. It is a security philosophy referring to adopting multi-scale and multi-modal defense measures for developing a robust security solution resulting in unattractive (costly) targets for potential adversaries. The underlying notion of a defense-in-depth model is utilizing the available resources in an organization to implement multi-layered protective and detective defenses to restrain adversaries from attaining their malicious objectives [33]. Section 3.3 briefly discusses the cyber-physical defense-in-depth model, which can be used as a benchmark to evaluate the defense status quo of a manufacturing system and identify if potential defense layers are missing.

### 3.3 Cyber-physical defense-in-depth model for manufacturing systems

The defense-in-depth security model relies on multiple overlapping and complementary security measures where each layer is designed to address different threats and vulnerabilities. Section 3.3.1 presents an overview of the traditional defense-in-depth model used in the IT industry, and the cyber-physical defense-in-depth model for manufacturing is introduced in Section 3.3.2.

#### *3.3.1 Traditional defense-in-depth model*

In traditional cyber-domain security, the defense-in-depth strategy involves implementing defenses within each layer of the cyber architecture of an enterprise, presented in Figure 3 (left). For example, the perimeter security layer protects information assets against attacks and threats [46], the network security layer defends against known network attacks through firewalls, intrusion prevention systems, and intrusion detection systems [47], the host security layer includes antivirus, host intrusion detection system, host-based firewalls, and operating system hardening [46], session security ensures web security via encrypted keys and session identifiers [48], and the application security layer protects the user input and information, validates provided inputs, and supports access control. This defense model also includes physical defenses in the form of a perimeter protection layer for server rooms, internet cables, computers, physical barriers, gates, security guards, lighting, locked facilities, surveillance cameras, access cards, and perimeter intrusion detection [33,49]. However, additional non-cyber defense measures summarized in Table 1 are recommended to realize the defense-in-depth strategy for manufacturing systems. It is worth mentioning that the NIST cybersecurity framework [50] only provides a foundational overview of potential defenses without providing a comprehensive characterization, their operational principles, and contextual relevance within the broader cybersecurity landscape.

#### *3.3.2 Cyber-physical defense-in-depth model*

Policies and procedures encompass comprehensive security strategies that define the organization's security posture, goals, and responsibilities. This forms the foundational layer of the defense model and establishes the framework and guidelines for all subsequent cyber and physical defense layers. All cyber domain security measures can be integrated into the cyber defense layer presented in Figure 3 (right), and additional non-cyber detective and protective defenses can be developed for manufacturing systems by leveraging the available physical resources in an organization. Hence, attention has been given to the available Quality Control (QC) resources on which manufacturing systems have relied for ages. Shafae et al. (2019) demonstrated that quality control regimes could be leveraged for developing detective non-cyber defense layers for manufacturing systems if designed with security in mind [15].

The first category of QC resources that can be developed into a non-cyber detection layer is personnel. With proper training and experience, personnel can detect physical changes to a manufactured part resulting from C2P attacks and raise alarms. Another potential QC resource is the widely used inspection tools and techniques that can be re-designed to detect attacks on manufacturing assets. Given a specific manufacturing application, these tools can be developed to detect the physical manifestation of cyberattacks in manufactured parts during post-production inspection. For instance, commonly used inspection tools such as the coordinate measuring machine (CMM) are programmed to check specific features (e.g., number of holes, the dimension of a hole), often known as the key quality characteristics (KQCs), in a manufactured part for post-production inspection. DeSmit et al. (2017) presented that the CMM will fail to detect any alterations in part geometry resulting from a cyber-physical attack that does not affect the features the CMM was programmed to check [1]. For example, adversaries with prior information about the inspection procedures can launch C2P attacks by tampering with CAD files. One solution approach can be the inspection of non-KQC features, using a 3D scanner that can capture numerous features simultaneously and compare these with the base model. Finally, real-time process monitoring to automatically identify, diagnose, and counteract any anomalies in a product/process is another QC regime that can be improvised for extending non-cyber defense layers [51]. In-situ sensor measurements of different process variables (e.g., temperature, cycle time) can be employed to monitor Key Performance Indicators of various processes so that anomalies within a control system can be detected, thus offering another venue for defending against cyber-attacks [52]. If the traditional cyber defenses fail, trained personnel, security-aware inspection methods, and security-aware processes will provide additional avenues to detect the physical manifestation of attacks on manufacturing systems.

**Table 1** Recommended defense layers for cyber-physical defense-in-depth

| References | Recommended defense layers | | | | |
|---|---|---|---|---|---|
| | Policy and procedure | Cyber | Personnel | Inspection | Process |
| NIST CSF 2.0 [50] | ✓ | ✓ | ✓ | ✓ | ✓ |
| Renaud et al. (2024) [53] | | | ✓ | | |
| Rahman et al. (2023) [54] | ✓ | ✓ | ✓ | ✓ | ✓ |
| Kayan et al. (2022) [55] | ✓ | ✓ | ✓ | | |
| Mullet et al. (2021) [56] | | ✓ | | | |
| Mahesh et al. (2021) [57] | | ✓ | | ✓ | ✓ |
| Elhabashy et al. (2020) [21] | | | | ✓ | |
| Shafae et al. (2019) [15] | | | ✓ | ✓ | ✓ |
| Proposed cyber-physical defense-in-depth | ✓ | ✓ | ✓ | ✓ | ✓ |

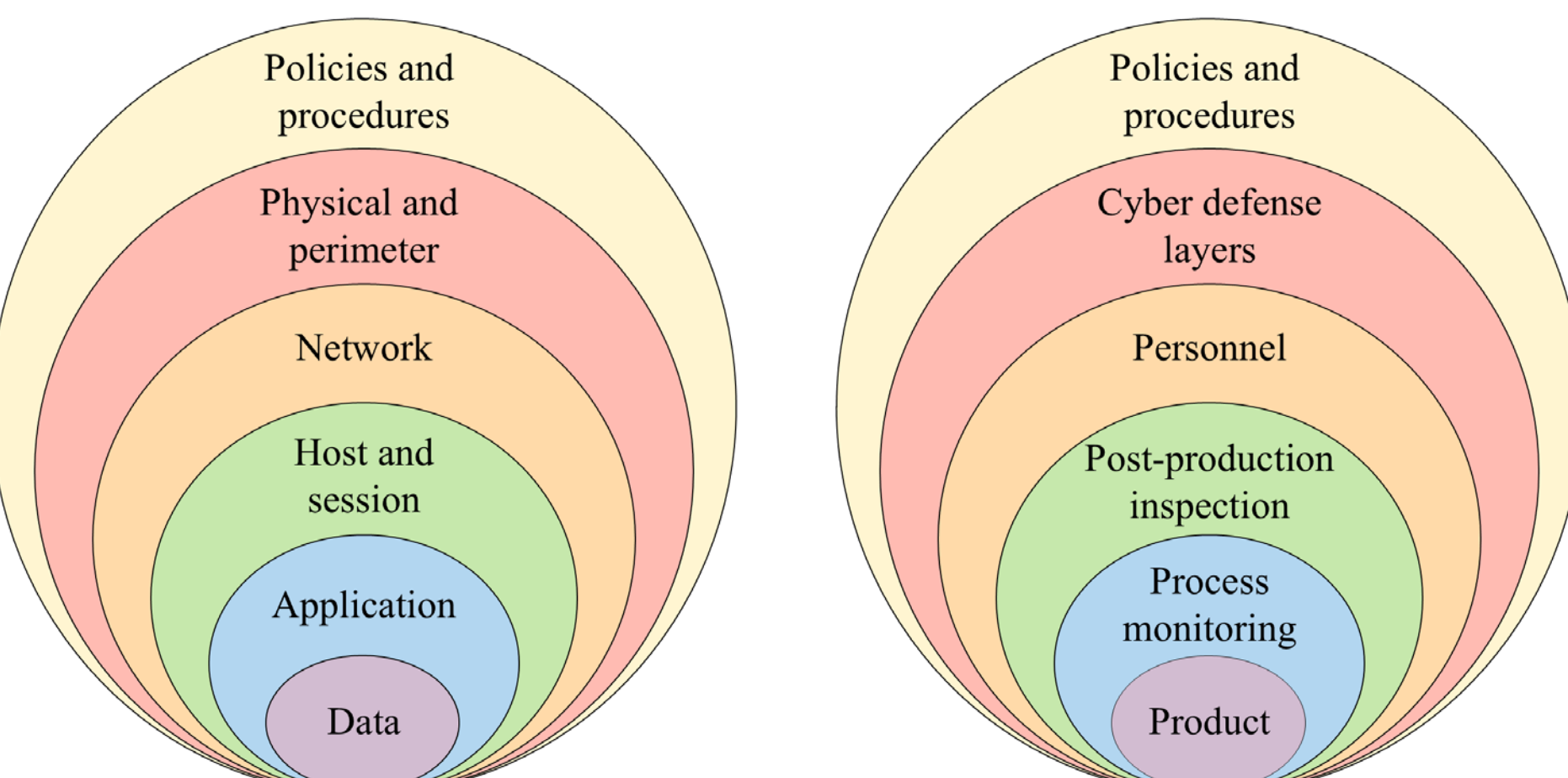

**Figure 3** Traditional defense-in-depth model (left) and cyber-physical defense-in-depth model for manufacturing systems (right)

In addition to the detective defense layers discussed above, non-cyber protective defenses can also be developed utilizing the knowledge of the physical aspects of production operations. Physical marking techniques such as the use of physically unclonable functions (PUFs) and fragile watermarks that are designed to break if tampered with, as well as embedding of internal identification codes (e.g., a QR code), can be effective defenses against theft of IP, counterfeiting, and illegal reverse engineering [58,59]. Additional security measures can be embedded as specific design features in the CAD file. For example, security features can be developed via design elements like curvatures, scaling functions, and overlapping surfaces using a specific combination of slicing operations and manufacturing processing parameters to obscure the design file for adversaries [60]. A part manufactured from an altered design file containing such security features will be distinct in appearance compared to the on-screen representation of the geometry. Such protective measures can be embedded into organizational policies, personnel (e.g., training), inspection, and process defense layers. All non-cyber detective and protective defenses combined with the cyber defense layers constitute the cyber-physical defense-in-depth model for manufacturing systems.

### 3.4 Cyber-physical defense-in-depth model-driven vulnerability identification

The cyber-physical defense-in-depth model provides a systematic framework to identify, audit, and assess vulnerabilities. Systems vulnerabilities can appear if (1) potential defense layers are missing and/or (2) defense measures are inappropriately designed and implemented. Therefore, the proposed vulnerability identification approach is hierarchical, which assesses the security landscape for each layer and identifies vulnerabilities across them. First, any missing defense layers from the cyber-physical defense-in-depth model can significantly increase cyber-physical vulnerabilities in manufacturing systems and expose the system to various attack vectors. For example, the inspection tools and techniques may not be designed as a physical defense layer to detect attack-induced alterations like geometric and dimensional changes in a product [6]. Commonly used post-production quality inspection and control methods are designed based on specific assumptions (e.g., a sustained shift in the process) and decision-making rationale (e.g., observing only a few key quality characteristics) that may become invalidated by cyberattacks. In such cases, a significant vulnerability for a manufacturing system is the absence of a security-aware inspection system. Second, even when the necessary defense layers are in place, their inappropriate implementation, misconfiguration, and insufficient integration can render defenses ineffective, which will have inherent vulnerabilities that adversaries can exploit. Each defense layer can be assessed independently with diverse toolsets and methods. For example, penetration testing can help identify equipment and manufacturing process vulnerabilities, whereas vulnerability scanning tools can focus on network vulnerabilities. In this work, relevant literature from Web of Science, Scopus, IEEE digital library, ScienceDirect, and ASME Digital Collection databases and conventional vulnerability repositories (e.g., the National Vulnerability Database (NVD) and Common Weakness Enumeration (CWE) [30,61]) were surveyed to characterize, identify, and classify potential vulnerabilities within each layer of the cyber-physical defense-in-depth model. Potential vulnerabilities in different defense layers are summarized in Section 4.

## 4 Cyber-physical vulnerabilities in manufacturing systems

This section provides a structured classification scheme to characterize and classify vulnerabilities in the five defense layers of the cyber-physical defense-in-depth model. The primary classification of cyber-physical vulnerabilities in manufacturing systems is depicted in Figure 4. The following subsections explain how each defense layer can become vulnerable, categorize vulnerabilities, and present individual vulnerabilities within each category.

### 4.1 Policies and procedures vulnerabilities

Policies and procedures can serve as a defense layer against cyberattacks by establishing structured guidelines for secure operations, risk management, regular security evaluations, contingency planning, and incident responses within an organization. While designed to enhance the security posture of smart manufacturing systems, organizational policies and procedures can inadvertently introduce vulnerabilities due to several factors. First, the lack of thorough and tailored cybersecurity policies leaves manufacturing systems vulnerable to the evolving cyberattack landscape. All security frameworks, including the NIST Cybersecurity Framework Manufacturing Profile [62], primarily focus on cyber-domain security, concentrating on data and software security. In contrast, physical domain security is not well-defined and limited to “physical access control”. Second, the guidelines presented in security frameworks are mostly generic, and effectively managing the cybersecurity risk requires a clear understanding of the business drivers and security considerations specific to the manufacturing system and its environment. Consequently, generic security policies are poorly understood, inadequately communicated, and improperly enforced, leading to critical security gaps in manufacturing organizations.

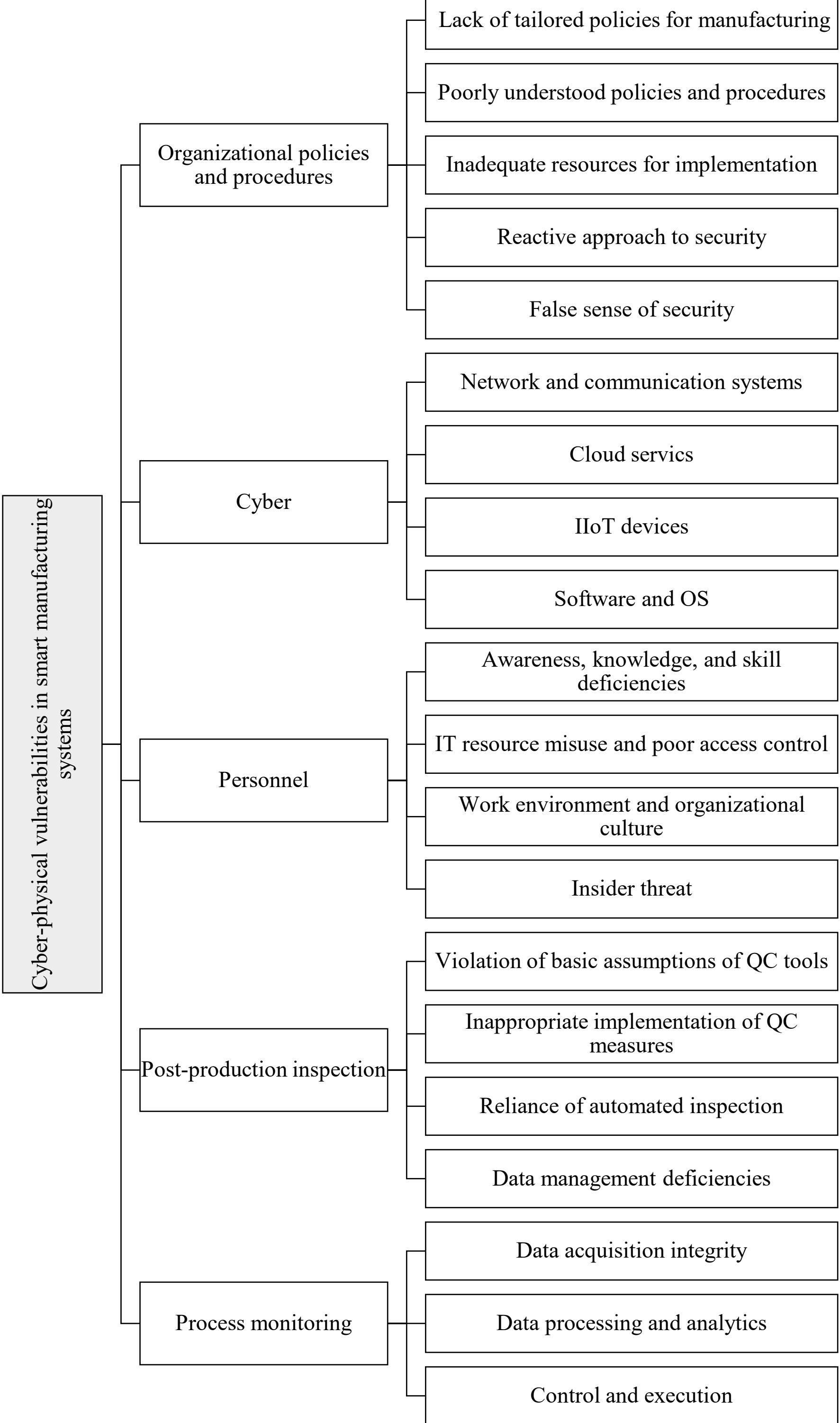

**Figure 4** Primary categories of cyber-physical vulnerabilities in manufacturing systems

Third, inadequate resources—especially in small and medium-sized businesses—hinder the effective implementation of rigorous security policies [63]. Organizations often try to circumvent security regulations to make operations cost-effective and incidentally introduce new security vulnerabilities [4]. Industry reports show that cybersecurity teams are short-stuffed and significantly underfunded in around 70% of manufacturing companies [64]. Fourth, the traditional reactive approach to cybersecurity in manufacturing means that threats are tackled mainly after major security breaches, allowing adversaries to exploit existing vulnerabilities and cause significant damage before any defensive measures are implemented [65]. Fifth, manufacturing corporations often create a false sense of security from their overconfidence in threat detection. According to industry reports, manufacturers are more confident in their cybersecurity preparedness than in their ability to respond to and recover from cyberattacks [66]. However, most organizations are unaware of the types and extent of emerging security threats when using various IoT devices and advanced digital technologies, especially in the smart manufacturing environment.

## 4.2 Manufacturing cyber domain vulnerabilities

This section presents vulnerabilities within the enabling tools and technologies in the cyber domain of smart manufacturing systems. While integrating IT technologies provides increased interoperability and better control in physical manufacturing systems, the operational technology assets have become cyber-accessible and a part of the growing and diverse attack surface [24]. Figure 5 presents the potential manufacturing cyber domain vulnerabilities explained in the following sub-sections. In addition to traditional cyber domain vulnerabilities, AI-driven cyber-attacks are emerging as a significant concern. For example, threat actors can leverage adversarial machine learning to manipulate and/or deceive intrusion detection systems or automate large-scale exploitation of system vulnerabilities, making cyber intrusions more sophisticated and challenging to detect [67].

### *4.2.1 Network and communication system vulnerabilities*

The network communication system in smart manufacturing systems can be vulnerable because of the used communication protocol, insecure data transferring/sharing, insufficient authentication and authorization, lack of encryption, and usage of removable media devices, as summarized in Figure 5. Vulnerable network communication systems may allow adversaries access to critical control systems and/or physical machines connected to the same network [68,69]. Threat actors can also send spoofed data traffic to different network resources to achieve their malicious objectives through man-in-the-middle attacks [70]. After accessing the network, they can steal confidential intellectual property about products and processes. Wells et al. (2014) and Sturm et al. (2014) demonstrated that adversaries could leverage network communication vulnerabilities for altering design files and tool path files along with numerical control commands/program files (e.g., G-code), leading to the production of defective products [20,71].

**Communication protocol**. Manufacturing plants commonly use distributed Local Area Network (LAN)[3] connections for integrated data acquisition and supervisory control. Plant networks usually have multiple entry points for streamlining different operations, often using outdated, inherently insecure protocols like FTP and Telnet [33,72]. Threat actors can target and exploit insecure communication protocols to intercept, manipulate, and disrupt critical data transmission [4,69,72–74].

**Insecure data transfer**. Insecure data transfer increases security risk by allowing attackers to intercept, alter, and steal sensitive information during data transmission [14,20,75–77]. An attacker may spoof and inspect data traffic in plaintext format and reverse engineer necessary unique protocols for obtaining authority over control communications between manufacturing assets [70].

**Insufficient authentication and authorization**. It can allow unauthorized users to access sensitive manufacturing equipment and data [20,69,73,76,78]. In manufacturing organizations, network services are usually run with default

[3] Local Area Network (LAN) is a common communications link to a server shared by a group of computers and associated devices

security configurations, inadequate firewalls, and sub-standard to non-existing authentication measures. Consequently, many physical ports are left open, and executable codes remain accessible, which adversaries may exploit [33]. Threat actors can also access control functionalities in a manufacturing system through discovered network backdoors[4], as the communication system is often deployed without sufficient security analysis.

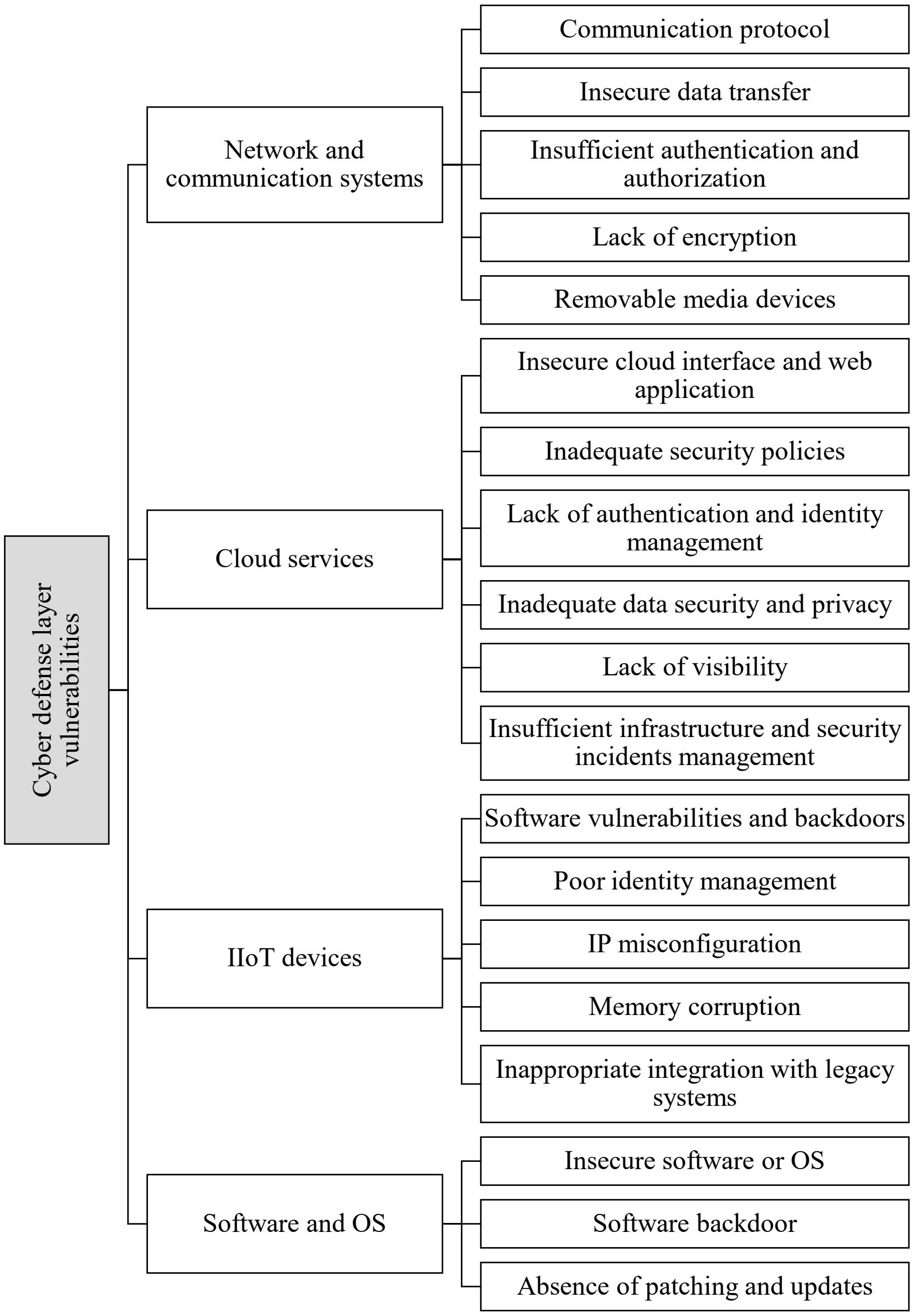

**Figure 5** Taxonomical classification of vulnerabilities in the manufacturing cyber defenses

**Lack of encryption.** Lack of encryption raises security risks by leaving data vulnerable to interception and unauthorized access during transmission, leading to potential data breaches and exploitation [73,78].

**Removable media device**. Removable media devices can introduce malware, facilitate unauthorized data exfiltration, and allow tampering with critical system files [14,73,78–81]. In recent years, the increased usage of virtual personal assistant (VPA) and voice-controlled smart assistant devices and services has also empowered adversaries to tap into employees' personal devices that can later be used as an entry point to the networks to which those devices are connected [82–86].

[4] Backdoors are deficiencies in the network architecture, or embedded potentialities that are forgotten, overlooked, or simply disregarded

#### *4.2.2 Cloud services vulnerabilities*

Insecure cloud interfaces and web applications, inappropriate security policies, lack of authentication and identity management, inadequate data security and privacy, lack of visibility, and insufficient infrastructure and security incidents management are the primary vulnerabilities in cloud services used in manufacturing organizations. While cloud services are still evolving, these inherent vulnerabilities are captivating targets for adversaries.

**Insecure cloud interface and web application**. Cloud service providers offer customers different software interfaces for connecting to their services. Highly user-friendly interfaces often have weak security measures and may reveal various flaws in security issues [87,88].

**Inadequate security policies**. As most cloud services are public, anyone with malicious intent can subscribe to a cloud service to study the critical vulnerabilities for exploitation [89]. Confidential cloud data can also be exposed to adversaries for not being cached properly. For instance, First American Corporation, a giant real estate and insurance company, mistakenly exposed 885 million sensitive financial records of their customers that were accessible to anyone on the company's website [90].

**Lack of authentication and identity management**. Threat actors can infiltrate and manipulate critical production and product data without proper authentication and identity management in cloud services, resulting in intellectual property theft and potential production disruption [89,91].

**Inadequate data security and privacy**. Data security and privacy are also among the major concerns in cloud services [92–97]. Failure to ensure data confidentiality, integrity, and availability may result in IP theft, production disruption, and compromised product quality and reliability. For example, design files (e.g., CAD models) are often shared over or saved in cloud storage. Threat actors can covertly steal or alter such files by exploiting the vulnerabilities of cloud storage [77]. Stolen information can be used by adversaries to produce counterfeit products, as well as to launch coordinated cross-domain attacks in the future [15].

**Lack of visibility**. The lack of visibility in cloud services can reduce the ability to monitor and detect unauthorized access, misconfigurations, and data anomalies, increasing the risk of cyberattacks and data breaches [65].

**Insufficient infrastructure and security incidents management**. It can allow threat actors to exploit vulnerabilities in cloud-based applications and services, compromising sensitive data [98,99].

#### *4.2.3 IIoT device vulnerabilities*

IIoT devices can be attacked physically (e.g., node tampering) and via cyber domains (e.g., traffic analysis attack). Common vulnerabilities in IIoT devices include software vulnerabilities and backdoors, poor identity management, IP misconfiguration, memory corruption, and inappropriate integration with the legacy system. It is worth mentioning that embedded sensors and electronics are now widely used in the manufacturing industry automation and control for collecting data, observing different production operations, and monitoring process parameters [100]. The data collected from the IIoT devices are used for various decision-making activities, such as activating specific valves in a hydraulic machine. Consequentially, compromised devices may fail to collect reliable data and/or relay fake data, leading to erroneous decision-making [101].

**Software vulnerabilities and backdoors**. Software vulnerabilities and backdoors allow adversaries to gain control of the network layer of these devices [88,98,102]. For example, a temperature sensor used in a production facility may have malware installed in it, which can be activated upon installing that sensor. Such a compromised sensor can send sensitive data packets to adversaries and/or send fake data to the controller.

**Poor identity management**. Adversaries can also exploit poor identity management practices (such as using a common username and password for controlling devices) to gain access to on-field active IIoT devices [91,103].

**IP misconfiguration**. IP misconfiguration can expose IIoT devices to unauthorized access, network attacks, and data breaches due to incorrect or inconsistent network settings [98].

**Memory corruption**. Memory corruption in IIoT devices arises when an attacker leverages software weaknesses to manipulate or corrupt the memory, leading to unauthorized code execution, system crashes, or takeover of device control [98].

**Inappropriate integration with legacy systems**. Integrating IIoT devices with legacy manufacturing equipment can create insecure web interfaces and interoperability issues that threat actors can exploit [99].

Additionally, software algorithms used in IIoT devices can be tampered with, leading to misinterpretation of the perceived signal and initiating wrong actuation commands [101]. However, potential risks and effects of such vulnerabilities on corporate networks are generally disregarded when installing IIoT-enabled devices [104].

#### *4.2.4 Software and OS vulnerabilities*

Insecure software or OS, software backdoors, lack of input validation, absence of patching and updates, and poor coding practices are the critical vulnerabilities in software and OS that pose a severe threat to manufacturing organizations firms that rely on myriads of software packages starting from design software (such as SolidWorks, and CATIA) to Enterprise Resource Planning (ERP) and Product Life-cycle Management (PLM) software. Production processes mostly start with the product design phase.

**Insecure software and OS**. Critical vulnerabilities, such as stack-based buffer overflows, have been found in product design software packages widely used in manufacturing companies [105,106]. Similarly, cybersecurity concerns are overlooked in PLM software, which contains highly sensitive information such as product specifications, production process plans, and intended product usage across the service and disposal of a product. Adversaries can target the above software to access confidential details, allowing them to launch coordinated cross-domain cyberattacks. For example, knowing the inspection procedure of an organization will enable an attacker to develop an attack scheme that can bypass quality inspections [15]. Specific knowledge of manufacturing systems, especially manufacturing control, can also be leveraged to attack and disrupt an organization's software processes. For example, security firms have discovered "EKANS" ransomware supposedly designed to specifically target software used in industrial control systems [107].

**Software backdoor**. It can enable hidden entry points for unauthorized users, allowing threat actors to evade typical security procedures and even obtain control over the system [108]. Software often has backdoors intended for the maintenance and upkeeping of software or systems. Usually, backdoors are kept for administrative purposes, only known to the software developer, and safeguarded with a hardcoded username and password that cannot be altered. However, threat actors can exploit such backdoors to access a system or data [109]. Backdoors create a portal for bypassing a closed system's encryption and authentication measures, which an insider can also create intentionally to assist adversaries in infiltrating the system [110].

**Absence of patching and updates**. The absence of patching and updates in software and OS poses a severe security threat since it can expose the system to known exploits and vulnerabilities [20,71–73,81]. Many computer systems in manufacturing companies operate with archaic software and OS versions (such as Windows XP and Windows 7) with no security updates and patches [111]. Even when software updates are available, manufacturing systems can still be vulnerable because the software cannot be updated and patched regularly to prevent production disruption, leaving the system exposed to attackers.

### 4.3 Personnel defense layer vulnerabilities

Personnel refers to anyone who can interact with manufacturing assets, from shop floor operators, machinists, mechanics, maintenance personnel, and shipping and material handling personnel to manufacturing engineers, quality engineers, designers, IT support staff, and visitors [1,15]. In smart manufacturing systems, personnel can access physical equipment to software tools such as the human-machine interface, supervisory control and data acquisition system, manufacturing execution systems, and manufacturing intelligence. Hence, insiders pose a significant security threat to manufacturing organizations with intimate knowledge of the system's operation, often unrestricted access to critical equipment and processes, and the ability to exploit vulnerabilities within the organization. Security can also be compromised by personnel without malicious intent from mishandling sensitive information, being unaware of the cybersecurity threats and their manifestation on the production floor, and not following proper security protocols. Manufacturing organizations also overlook the need for adequate cybersecurity training and employee awareness, making personnel the most vulnerable node in the system [112,113]. Figure 6 shows the potential categories of vulnerability that can exist in the personnel defense layer, which are often correlated. It is worth noting that recent advancements in AI have enabled deepfake phishing attacks, where synthetic media is used to impersonate trusted individuals, deceiving employees into granting unauthorized access to critical systems. These AI-driven social engineering tactics significantly amplify the risks posed by human factor vulnerabilities [114,115].

#### *4.3.1 Awareness, knowledge, and skill deficiencies*

Awareness, knowledge, and skill gaps are some of the crucial vulnerabilities in personnel that result in a lack of preparedness against evolving cyber threats and increasing the risk of operational disruptions.

**Limited cognitive sensitivity**. Limited cognitive sensitivity of employees refers to the limited experience/knowledge of a product or process, which can lead to failure in detecting any alterations in that product or process resulting from cyberattacks. The personnel may have limited cognitive sensitivity to the (1) product, (2) process, and (3) equipment and technology [15]. In a highly automated machine shop, for example, the machine operators' roles shift to multiple parts loading, setup, and unloading on multiple machines, limiting their cognitive knowledge of the product specifications. Hence, they may be unable to identify small but impactful malicious geometric and dimensional changes in a product.

**Lack of cybersecurity awareness and training**. Personnel overlook cyberattacks as possible failure modes without cybersecurity awareness and proper training. In two case studies, Wells et al. (2014) and Sturm et al. (2014) demonstrated the possibility of tampering with a product's tool path file and STL file, respectively, producing parts with incorrect dimensions. In both studies, participants failed to identify the cause of producing incorrect parts, and they were convinced that the problem was caused by some error in the respective production processes [20,71]. Personnel who lack cybersecurity awareness and proper training are also prone to phishing emails, malicious links, and social engineering attacks, becoming easy targets for adversaries [20,73,116]. Smart manufacturing systems comprise numerous interconnected IoT devices and digital technologies, creating a broader attack surface. Lack of awareness and training can lead to improper handling of devices and technologies because of their insufficient understanding of the associated cybersecurity risk.

**Shortage of trained cyber-physical security professionals**. The shortage of trained security professionals significantly weakens a manufacturer's ability to develop, implement, maintain, and monitor security protocols and policies tailored to cyber-physical manufacturing systems' unique cybersecurity threat landscape. For example, insufficient expertise may result in improperly configured industrial control systems, unsegmented networks for production systems, and delayed application of software patches in IoT devices, leaving production systems vulnerable to attacks.

#### *4.3.2 IT resource misuse and poor access control*

Inappropriate use of IT resources and poor access control can allow adversaries to exploit compromised devices and credentials to access manufacturing systems and execute attacks.

**Inappropriate use of IT resources**. Misuse of work devices for personal activities, connecting personal devices to the office network, and storing and sharing data on personal devices increase the risk of data leakage and exposure to potential malware. For instance, a factory-issued laptop used for browsing non-work-related websites can inadvertently download ransomware that propagates across the production network, or employees may lose their personal devices and media containing confidential data and leak an organization's critical data [116,117].

**Weak passwords**. Simplistic and reused passwords put systems at risk of unauthorized access. Additionally, most industrial control systems and operational technology devices come with default passwords that are usually left unchanged. For example, researchers found critical password-related vulnerabilities in an industrial serial-to-Ethernet converter, which contained a hardcoded root username and password combination that could be easily extracted and couldn't be changed [118]. Some programmable logic controllers also inadvertently expose the password, allowing unauthenticated adversaries to remotely obtain the device's password in plain text [119].

**Weak authentication practices**. Most manufacturing systems lack multi-factor authentication, which can allow unauthorized remote access and increase the risk of credential misuse. For example, adversaries can gain remote access to programmable logic controllers by exploiting the single-factor authentication and reconfiguring machinery, leading to costly downtime and quality issues. Recent studies also reveal that 44% of manufacturers have sensitive files accessible to all employees, indicating that a single compromised account can result in extensive intellectual property theft [120].

**Neglect of security best practices**. Manufacturers often neglect to implement and maintain security best practices, such as updating software regularly, leaving known vulnerabilities unaddressed. According to an industry report, 60% of data breaches were caused by unpatched known vulnerabilities [121].

#### *4.3.3 Work environment and organizational culture*

An organization's work environment and cultural attitudes toward cybersecurity significantly impact the overall security posture, and the lack of leadership support and poor communication are major barriers to securing manufacturing systems.

**Lack of leadership support**. Lack of leadership support and insufficient emphasis on cybersecurity frequently results in inadequate resources allocated to cybersecurity initiatives and undermines adherence to security protocols. For example, if leadership fails to prioritize cybersecurity and mandate cybersecurity training, employees may not receive adequate training or tools to defend against threats. This can result in employees unintentionally introducing vulnerabilities, such as using weak passwords or falling for phishing scams.

**Absence of accountability**. The lack of accountability creates an environment where employees may not feel responsible for following security protocols. Critical vulnerabilities and security incidents may go unaddressed without clear ownership of security tasks. For example, suppose no one is specifically responsible for patch management for the CNC machines on a shop floor. In that case, systems may go unpatched for long periods, leaving them vulnerable to known exploits.

**Lack of incident reporting**. Employees may choose not to report security breaches due to unclear or strict policies. For example, an operator noticing unusual data trends in the production system may not report them for fear of being blamed or having their concerns dismissed, allowing the problem to go unnoticed. According to recent studies, 45% of enterprises encounter employees concealing cybersecurity incidents, potentially to avoid punishments [122].

**Fatigue and overwork**. Fatigue and overwork significantly impact an employee's ability to remain vigilant against cybersecurity threats. Employees who are overburdened, particularly those in cybersecurity roles, are more likely to experience burnout, which leads to low self-efficacy and skepticism toward extensive security procedures. Employees experiencing burnout often perceive cybersecurity measures as "not worth the hassle". Consequently, exhausted employees are more likely to make mistakes or disregard security protocols completely. Industry reports reveal that 95% of cybersecurity professionals are overworked, and nearly 70% of employees have bypassed security procedures [123].

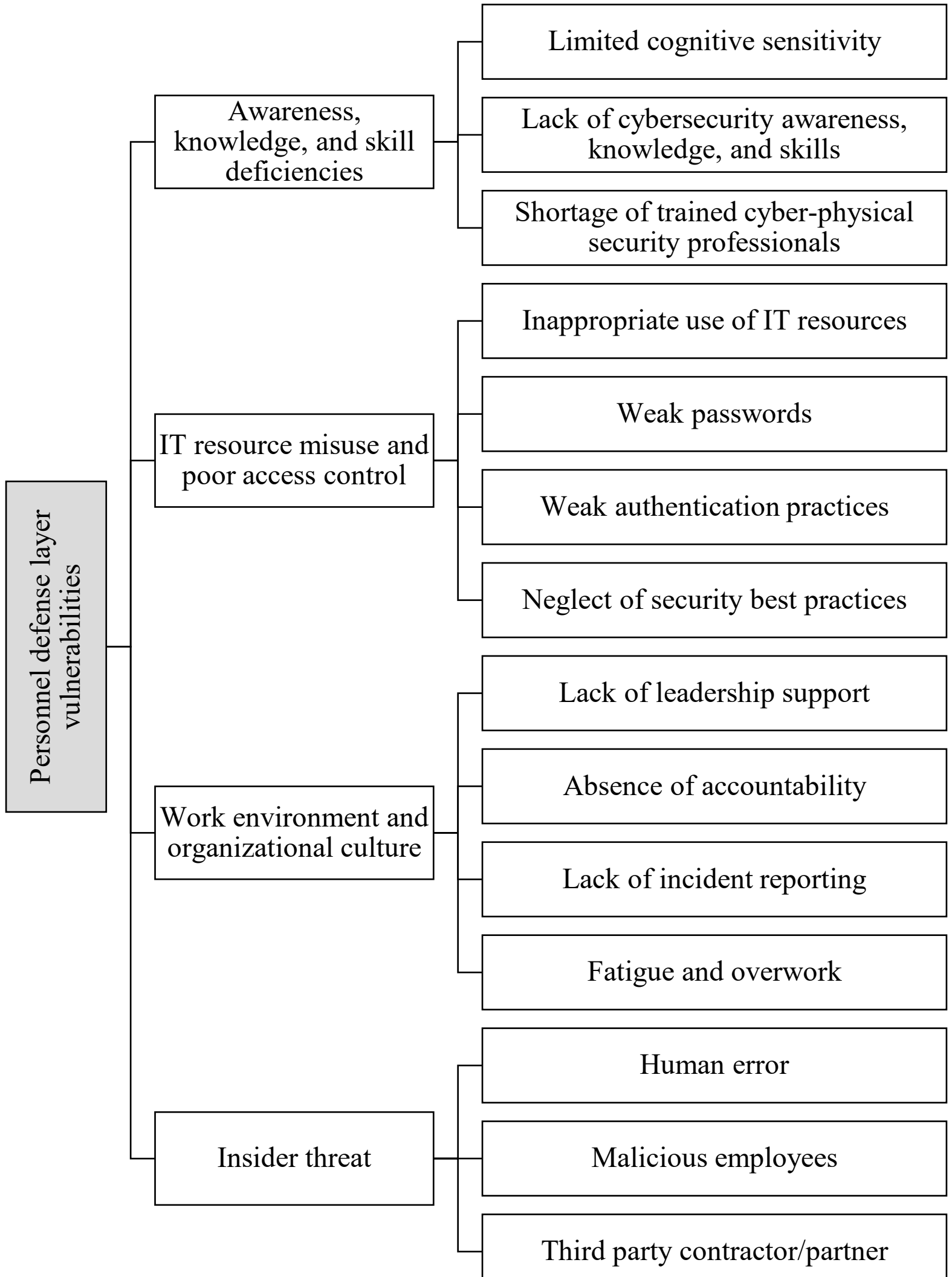

**Figure 6** Taxonomical classification of vulnerabilities in the personnel defense layer

### *4.3.4 Insider threat*

Insiders refer to authorized users with legitimate access to a company's assets and information who deliberately or accidentally abuse their access or privilege. Insiders can exploit their knowledge, privileged access to the organization's data, and the trust provided to them, creating a unique set of vulnerabilities for manufacturing systems.

**Human error**. Inadvertent mistakes or unintentional human errors stem from ignorance, lack of information, and misjudgment, which are common and leading causes of security breaches [124]. According to industry reports, 24% of the data breach incidents involved human error and/or negligence of employees, resulting in a total cost of 3.5 million US dollars [125]; insider threats annually cost organizations 11.45 million US dollars on average [126].

**Malicious employees**. Disgruntled and malicious employees can use insider access to harm an organization through espionage, unauthorized disclosure of critical information, and degradation or stealing organizational assets and capabilities [127]. Malicious insiders can utilize their access to confidential information and/or physical access to various equipment in the production plant to give away classified IP, disrupt operations, and even sabotage equipment by plugging in a malicious USB device, making slight changes in a product's design file, and changing production process parameters [128–131].

**Third-party contractor/partner**. Third-party contractors, subcontractors, suppliers, and vendors in the manufacturing supply chain have de facto insider access to various manufacturing systems' resources, such as product design and life cycle data. Integrating an external supplier's network and sharing data with third-party distributors creates security risks for manufacturers. Partners can intentionally leak sensitive data, change the integrity of supplied raw materials, and embed hardware backdoors into products and equipment [132].

### 4.4 Inspection defense layer vulnerabilities

Manufacturing systems have relied on various post-production inspection tools and methods to ensure outgoing product quality and reliability [133,134]. Traditional post-production quality inspection and control methods are designed based on certain assumptions and decision-making rationale that cyberattacks may invalidate [6,135]. Therefore, understanding and identifying potential vulnerabilities in these approaches is crucial to prevent adversaries from exploiting those and maintaining a safe and reliable production environment. Potential vulnerabilities in the inspection defense layer are summarized in Figure 7 and explained in the following sub-sections.

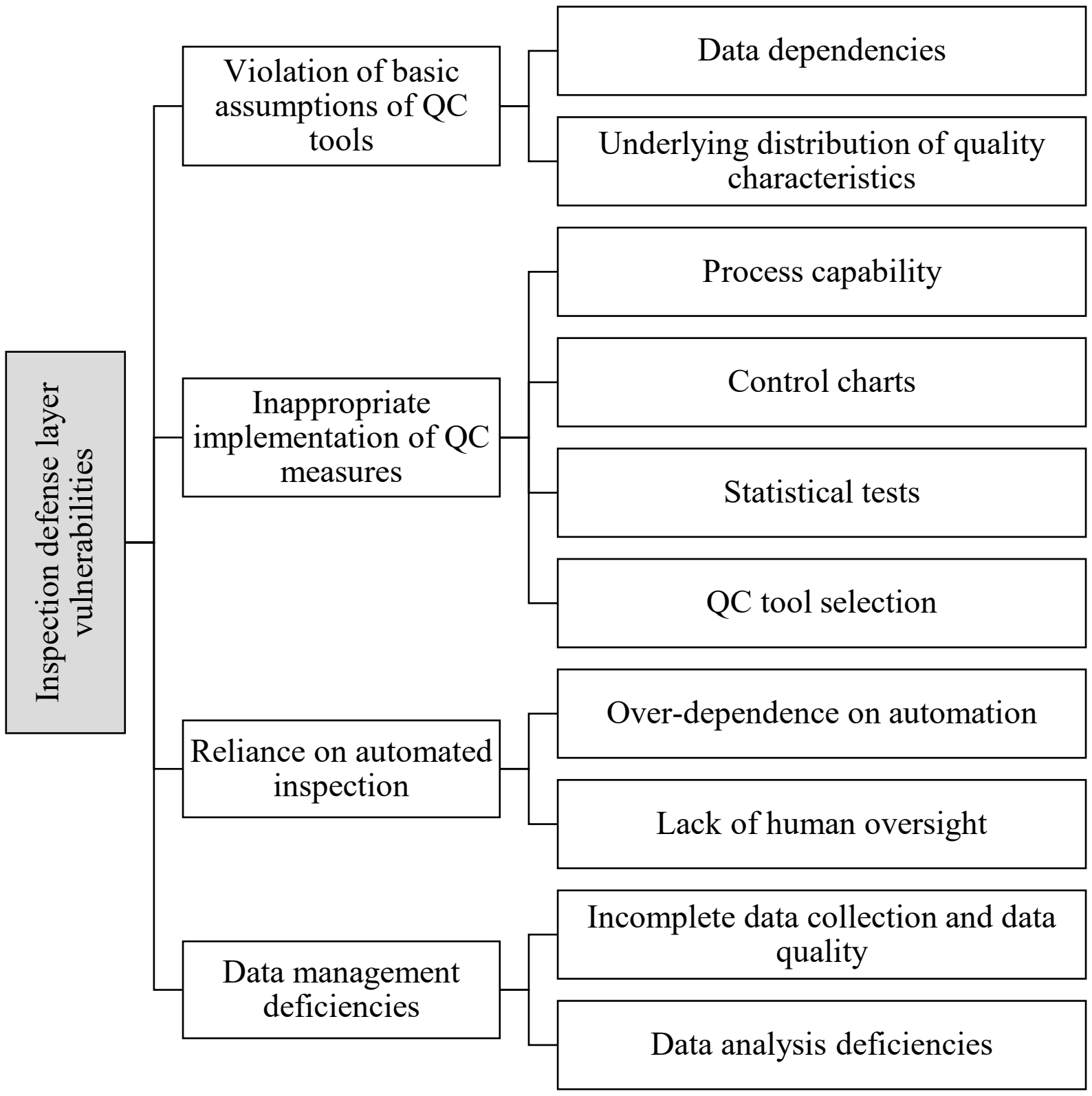

**Figure 7** Taxonomical classification of vulnerabilities in the inspection defense layer

#### *4.4.1 Violation of basic assumptions of QC tools*

QC tools rely on assumptions that can be inadvertently violated during implementation [20,134–137]. Specifically, overlooking data dependencies and incorrect assumptions about the underlying statistical distribution of quality attributes can compromise the sensitivity and reliability of the QC tools.

**Data dependencies**. The inspection data may be correlated across various production stages; treating such data as independent might conceal subtle patterns or trends that otherwise would point to a cyber-physical attack [21]. An attacker could exploit such oversight by implementing small coordinated changes to many process parameters, appearing within normal variation when viewed in isolation [15]. In additive manufacturing, for example, neglecting the interdependence between layer deposition sequences might lead to undetectable flaws should a cyber-attack target particular layers since the inspection process might not consider such correlations. Additionally, ignoring data dependencies may lead to the use of inappropriate statistical tools or control charts that assume independence, resulting in decreased sensitivity to detect out-of-control conditions caused by cyber-physical attacks.

**Underlying distribution of quality characteristics**. The effectiveness of QC tools in smart manufacturing systems depends on accurately characterizing the underlying statistical distributions of quality characteristics. Implementing QC tools by assuming incorrect data distributions can introduce vulnerabilities. For example, using control charts designed for normally distributed data on processes with non-normal distributions can result in misleading control limits, failure to detect anomalies, or false alarms. Such misalignments jeopardize the integrity of process monitoring, allowing deviations—possibly caused by cyber-physical attacks—to go undetected. As a result, a thorough understanding and validation of data distributions is required to ensure the effectiveness of QC tools in protecting manufacturing systems from both operational inconsistencies and malicious changes in products and processes due to cyber-physical attacks.

#### *4.4.2 Inappropriate implementation of QC measures*

Manufacturing organizations often implement QC tools without fully understanding the scope, implementation requirements, and significance of those tools. Inappropriate use of QC metrics, control charts, and statistical tests impedes reliable monitoring and introduces vulnerabilities that adversaries can exploit.

**Process capability**. Inappropriate implementation of process capability indices can cause failure to identify product and process-oriented cyber-physical attacks. Process capability quantifies how well a manufacturing process can consistently produce parts or products within specified tolerance limits by comparing the natural process variability to the design specifications. Process capability ratio, such as $C_p$, is commonly used to express the process capability quantitively. Process capability indices consider (a) the observed quality characteristic follows a normal distribution, (b) the process is in statistical control, and (c) the process mean is centered between the upper and lower specification limits [138]. Making inferences from the process capability indices without verifying data normality and/or statistical control of the process can deter the timely detection and identification of cyber-physical attacks by creating a false sense of process stability and quality.

**Control charts**. Misuse of control charts can also introduce vulnerabilities by undermining their ability to effectively monitor and control process variability. For example, inconsistent control limits, incorrect construction of control charts, misinterpretation of observed data points, overreacting to small shifts, and missing actual trends in the data can result in undetected process deviations or unnecessary interventions [21,137]. These issues can compromise the reliability of the inspection process and make it more vulnerable to cyber-physical attacks, in which malicious actors can exploit systemic flaws to evade detection [15]. For example, imposing control limits only on specific shifts (e.g., day shift only) creates monitoring gaps, leaving other shifts vulnerable to undetected anomalies [139]. Using erroneous limits derived from incorrect equations or limited data points and implementing tools suggested by management instead of selecting proper tools will impede the detection of assignable cause variations [140], allowing cyber-physical attack-induced physical changes to go undetected. The same control chart can also exhibit different sensitivities to varying process shift types [141], leading to failure in detecting various instantaneous and evolving process shifts [54]. Additionally, machine operators and shop floor managers often do not know how to read and interpret the widely used QC control charts, rendering the charts ineffective [136].

**Statistical tests**. Software packages for statistical tests[5] are often used without understanding why or how the tests should be conducted [142]. Due to inappropriate implementation, statistical process control tools or models can produce misleading results, and statistical methods may fail to reveal essential information [142]. Additionally, misinterpretation of the data and results and miscalculations can lead to frequent false alarms and unnecessary production disruptions.

---

[5] A statistical test quantifies evidence to support or reject a hypothesis about a process [175]. It evaluates whether sufficient data exists to reject the null hypothesis, which represents the assumed condition. Failure to reject the null may indicate either its validity or insufficient data to draw a definitive conclusion.

**QC tool selection**. QC tools can be misused because of their poor design and/or implementation, such as assuming that a specific tool applies to the observed quality characteristic. Selecting inappropriate QC tools can lead to misrepresentation and inferior performance. For example, using the exponentially weighted moving average (EWMA) chart to detect large shifts in the *process mean* may provide inaccurate results because EWMA is designed to detect small shifts [143].

#### *4.4.3 Reliance on automated inspection*

The growing reliance on automated inspection systems in manufacturing environments, while enhancing efficiency and precision, introduces new security concerns and vulnerabilities. This section describes the potential vulnerabilities arising from over-dependence on automated quality control processes and the lack of human oversight.

**Over-dependence on automation**. Increased dependency on automation and computer-aided support tools (such as CAD, CAM, and CAE[6]) increases the cybersecurity threat [14,73,143,144]. Manufacturers are increasingly automating post-production quality inspection through wide adoption and the use of IIoT devices, sensors, and cameras [143]. However, integration with the digital manufacturing network can put the QC system at risk of being compromised. For example, threat actors can launch passive joint attacks, collecting information about QC systems through PLM and using it to design attacks that can evade detection [143]. Vulnerabilities in support tools can also be exploited to bypass inspection. For instance, if the design file in machine vision systems is tampered with via CAD or CAM software, the attack may go unnoticed [73,144]. Note that CAD and/or CAM manipulation directly affects the geometric dimensioning and tolerancing (GD&T) information of the produced parts, which is the reference for inspection equipment. Therefore, if GD&T information is tampered with beforehand, inspection equipment will be useless down the production line. Additionally, sensors used for quality inspection can be tampered with to feed false data to the QC system, bypassing quality checks entirely.

**Lack of human oversight**. Automated inspection systems lack contextual understanding and adaptability inherent to human operators and inspectors. This makes them vulnerable to sophisticated cyber-physical attacks – introducing subtle and coordinated changes – designed to evade detection during post-production inspection. For example, adversaries may tamper with the design specification used for inspection, knowing that the absence of a human reduces the chances of anomalies being questioned or investigated. Besides this, without human intervention, automated systems may also miss out on adapting to evolvement strategies by attacking or picking up fine details that point to malicious activity, thereby opening up manufacturing processes and the quality of products to threat. Furthermore, increased dependence on automation may, over time, reduce human expertise and, therefore, further decrease the ability to identify or respond to unforeseen events. The lower human vigilance creates avenues through which cyber-physical attacks can make subtle changes undetected by automated systems, compromising product integrity and operational security. Human involvement will be necessary to ensure broad quality control and strengthen security measures against potential cyber threats.

#### *4.4.4 Data management deficiencies*

Inadequate data collection and analysis can lead to undetected changes in non-key quality characteristics, making the QC system vulnerable to product-oriented cyber-physical attacks [136,143].

**Incomplete data collection and data quality**. QC systems can become vulnerable when the collected data is insufficient or lacks the information needed to detect product-oriented cyber-physical attacks [21,145,146]. The vulnerability arises from (a) not collecting sufficient data for a specific feature, (b) collecting data for only a subset of features, and (c) inadequate data collection frequency. Adversaries can exploit knowledge of how manufacturers use the collected data for assessing the quality of a part feature to tamper with the product's geometric and functional integrity.

---

[6] Computer-Aided Engineering (CAE) tools are software applications used to simulate, analyze, and optimize engineering designs and manufacturing processes.

**Data analysis deficiencies**. In smart manufacturing systems, performing inadequate analysis and/or not exploiting data to the full might lead to losing crucial patterns or trends that could otherwise signal cyber-physical attacks. For example, the X-bar[7] control chart is often used in production shop floors in Phase II monitoring of products' KQCs without an accompanying Standard Deviation (S) or Range (R) chart. Using the X-bar chart alone without the R[8] or S[9] chart is an inadequate evaluation of the production process stability. The X-bar chart alone might fail to account for variability in the KQC introduced by intelligently designed cyber-physical attacks, whereas an accompanying R chart could help detect those [21].

### 4.5 Process defense layer vulnerabilities

Manufacturing process control can provide an additional defense layer against cyberattacks by continuously monitoring and regulating production processes to ensure they run within predefined specification limits. Self-adaptive systems with feedback controls can promptly detect deviations from the expected system behavior in production processes, create alerts and reports for diagnosis, and take mitigation actions. Figure 8 depicts such a system that maintains the desired characteristics of the process output $Y_p$, based on given input $U_i$. Process output $Y_p$ can fluctuate due to natural process variation, while it can also be affected by an external perturbation $d$. Hence, the system continuously monitors $Y_p$ using a sensor suite and feeds the sensor data $Y_f$ back to the controller. Next, the controller compares the user-defined reference $U_i$ with sensor data $Y_f$, and determines process correction $U_c$ for any observed system deviations from $U_i$ using pre-defined control algorithms. It also activates relevant actuators to perform an action $Z_a$ to maintain stable and desired system performance. However, anomaly detection techniques and self-adaptive systems are primarily developed to identify and respond to equipment and process failure without considering the potential for cybersecurity threats.

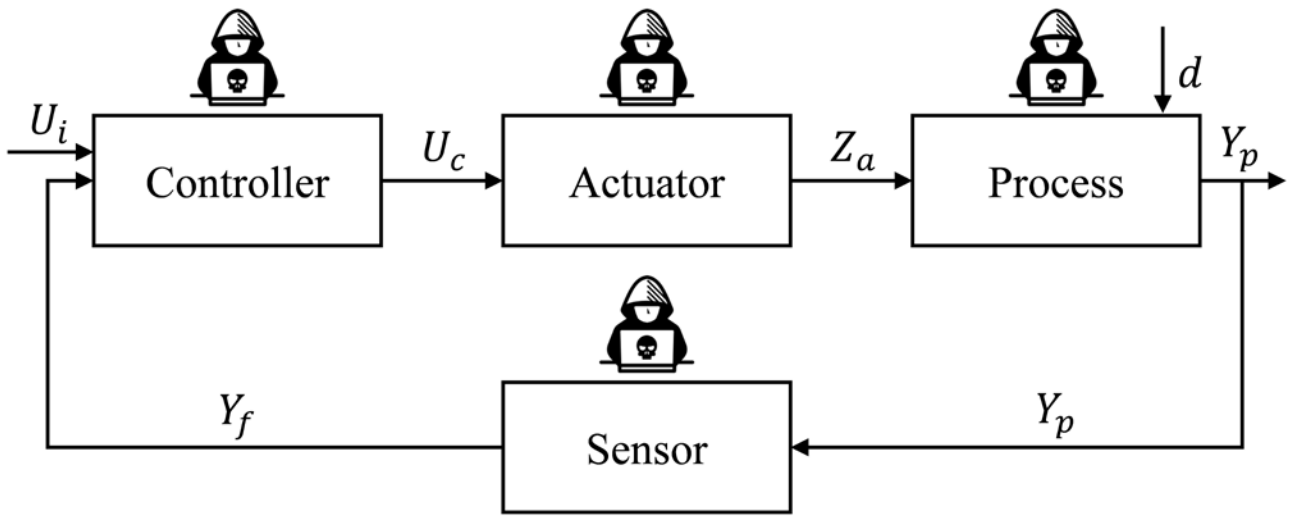

**Figure 8** Self-adaptive system with feedback control with vulnerable nodes

In manufacturing, adversaries can design cyber-physical attacks on the controllers, actuators, and sensors used in these systems and compromise the system's integrity [147]. Despite having guidelines from NIST about physical access control, many manufacturing companies, especially SMEs, cannot ensure proper control of physical access. The lack of security policies within organizations is often responsible for this. Physical access to a production facility allows for tampering with the equipment and hardware. Once adversaries have gained control over programmable logic controllers, actuators, or sensors, they can transmit forged control decisions to physical processes and/or report fake sensory data to the controller while complying with normal traffic patterns (e.g., connection logs) [148]. Attacks on the controller, actuator, and sensors will respectively tamper the correction $U_c$, actuation $Z_a$, and sensor inputs $Y_f$. If any elements of the feedback control are compromised, cyberattacks can easily go undetected. Additionally, adversaries can exploit physical access to any equipment to collect side-channel emissions (e.g., acoustics) for stealing IP or reconstructing the object [149–151].

---

[7] An X-bar chart monitors the stability of a process by tracking the average (mean) of sample measurements over time [176]. It consists of a center line representing the overall process mean and control limits that define the expected range of variation.

[8] An R (range) chart is used alongside the X-bar chart to monitor process variation within a set of samples [177]. It tracks the difference between the highest and lowest values in each sample group, helping to detect inconsistencies in process variability.

[9] An S (standard deviation) chart, similar to the R chart, measures process variation but uses the standard deviation of each sample instead of the range [178]. It is preferred for larger sample sizes because the standard deviation provides a more precise measure of variation.

Traditional approaches for monitoring machine tool conditions, process conditions, surface integrity, machine tool state, and chatter rely on sensor signal acquisition, data processing, feature extraction, and implementation of cognitive paradigms (e.g., fuzzy logic, genetic algorithm, and neural networks) [51]. Next, the execution and control step applies the corrective action determined using the cognitive paradigm via the actuators, thereby adjusting the system's behavior toward the desired setpoint. Therefore, understanding and identifying potential vulnerabilities in (a) data acquisition, (b) data processing and analytics, and (c) control and execution steps are crucial toward developing and implementing security-aware process monitoring in smart manufacturing systems. In response, this work proposes the first classification scheme for potential vulnerabilities in the process defense layer, summarized in Figure 9 and described in the following subsections.

#### *4.5.1 Data acquisition integrity*

Tampering with sensors and the calibration process can compromise the reliability and accuracy of sensor data, leading to erroneous decision-making and potential disruption of manufacturing processes. This section briefly describes potential vulnerabilities in the data acquisition step.

**Sensor data tampering**. Sensor data tampering directly compromises the reliability of the data used for monitoring and control. Physical tampering of sensors, such as damaging, misaligning, or obstructing sensors, can disrupt the measurement of key process variables, drive incorrect control actions, allow anomalies to go undetected, and leave the system prone to cascading failures. Additionally, sensor data interference can alter sensor signals to the control system [152,153]. Adversaries can distort sensor readings by injecting false signals and/or adding noise and trick the system into responding to fabricated conditions. For example, an attacker can inject low-temperature readings to decrease the use of coolant during machining, reducing product quality or causing damage to tools and equipment.

**Calibration errors**. Calibration error is the inaccuracy or inconsistency in sensor measurements resulting from incorrect or outdated sensor equipment calibration. Inaccurate sensor readings provide unreliable data to manufacturing control systems, resulting in incorrect process adjustments, undetected quality defects, or false alarms that disrupt production. Adversaries can leverage these vulnerabilities to introduce subtle product defects while evading detection. For example, an attacker with knowledge of sensor calibration status can design attacks that specifically target the error margins of poorly calibrated sensors, making their activities more difficult to distinguish from normal process variations.

#### *4.5.2 Data processing and analytics*

This section describes inherent vulnerabilities within the data processing and analytics phases of manufacturing process monitoring, highlighting how adversaries can exploit traditional feature selection and anomaly detection algorithms.

**Feature selection and extraction**. In situ monitoring techniques to detect conventional causes of variation (e.g., tool wear and chatter) and detecting machine states are primarily based on signal features selected to match specific detection targets (e.g., chatter onset). Such features may fail to identify product and process-oriented cyber-physical attacks [6]. Additionally, commonly used features like the mean power consumption, maximum vibration, average cutting force, and total time required for machining without considering the spatio-temporal nature of the signal can mask transient anomalies induced by attacks.

**Anomaly detection approaches**. Anomaly detection techniques and algorithms can inadvertently introduce vulnerabilities if not carefully implemented and regularly updated. First, the deployed algorithms may establish a limited characterization of "normal" behavior, potentially allowing adversaries to design attacks within the defined range of *normal behavior* [6,15]. For instance, adversaries could exploit the algorithm's learning period to gradually introduce malicious changes that become accepted as standard patterns. Second, anomaly detection algorithms often rely on historical data and predefined threshold values to identify anomalies [135], making them susceptible to adversarial attacks where malicious actors deliberately inject deceptive data to mimic normal behavior and evade detection. Third, the developed algorithms for manufacturing applications often lack robustness, i.e., the ability to perform reliably and accurately under varying conditions, including noisy, unexpected, or adversarial inputs [154]. Such algorithms can become vulnerable when they are overly sensitive to small perturbations in input data, lack adaptability to unseen scenarios, or are trained on biased or insufficient datasets, making them susceptible to errors, adversarial attacks, or exploitation by adversaries. Additionally, AI-enabled anomaly detection systems can be vulnerable to adversarial manipulation [155]. Threat actors can inject deceptive data patterns to either mask true anomalies or generate false alarms, undermining the reliability of automated monitoring frameworks.

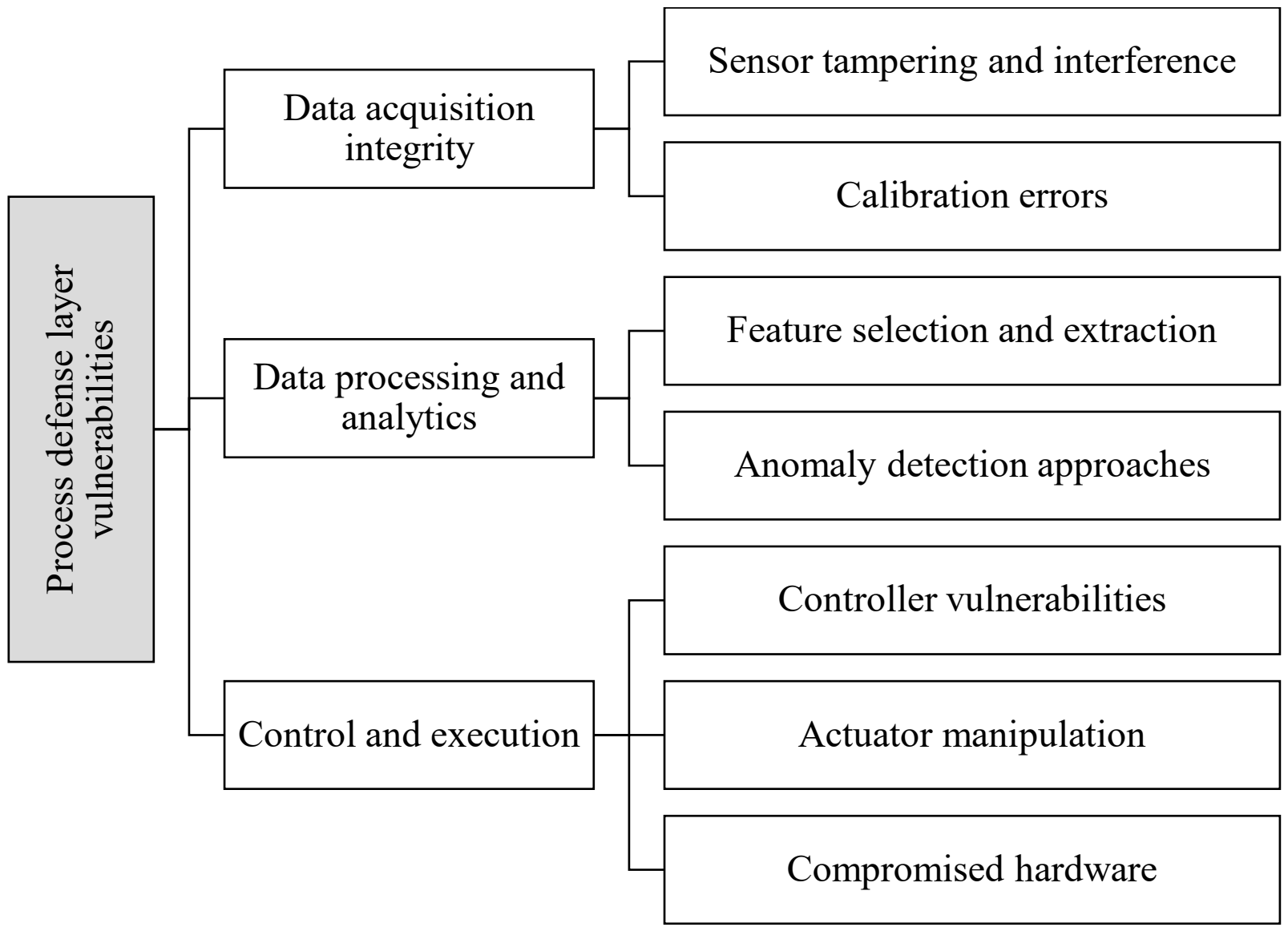

**Figure 9** Taxonomical classification of vulnerabilities in the process defense layer

### *4.5.3 Control and execution*

Control and execution vulnerabilities in smart manufacturing can arise from insecure controllers, manipulated actuators, and compromised hardware. These vulnerabilities enable attackers to infiltrate control systems, interfere with production processes, and compromise product quality and operational safety.

**Controller vulnerabilities**. Programmable logic controllers (PLCs) often lack integrated security features, leaving them vulnerable to cyberattacks. Due to the lack of anomaly detection or attack recovery mechanisms in PLCs, adversaries with basic knowledge of PLC control and command syntax can compromise PLC programs and memory [101]. Malware can also corrupt logic or algorithms within controllers. For example, the Siemens SPPA-T3000 distributed control system, despite its claims of secure operation, was reported to have significant vulnerabilities such as improper authentication, cleartext transmission of sensitive information, and unrestricted upload of digital files [156]. This highlights that threat actors can exploit even seemingly secure controllers remotely. Additionally, manufacturing control systems often use generic engineering models, tools, and techniques that are widely known and accessible, making it easier for adversaries to identify system vulnerabilities [157,158]. Successful exploitation of these vulnerabilities can enable attackers to execute arbitrary malicious commands, gain root privileges, access confidential information, and manipulate production process parameters [74].

**Actuator manipulation**. Modern manufacturing operations are controlled by various actuators that execute commands from programmable controllers [159], where sensor data is processed and analyzed using different algorithms for decision-making [101]. The signal an actuator perceives can also be changed, causing it to implement wrong decisions [160,161]. For example, in a beverage production line, sensors determine the level of filling in bottles, and a robotic end-effector pushes overfilled and underfilled bottles away from the conveyor belt. Potential attacks include the time delay attack that can shift the timer for activating the actuator by a few seconds so that a defective bottle will pass over the conveyor, and the actuator will push away a correctly filled bottle.

**Compromised hardware**. Compromised hardware can pose significant security risks to smart manufacturing systems due to poor password practices and firmware tampering. Most legacy machines have hard-coded passwords and login credentials (e.g., ID: *admin* and password: *12345*), making manufacturing an easy target for adversaries. Adversaries can tamper with machine hardware and firmware, leading to unwanted operations and making equipment unusable [63,77,81]. For example, Slaughter et al. (2017) demonstrated that adversaries could tamper with parts' geometric integrity by attacking the infrared imaging system (used for closed-loop QC) in powder bed fusion additive manufacturing [162]. Additionally, adversaries, especially state-sponsored threat actors, can embed unauthorized commands or codes into the firmware to remotely take control of manufacturing equipment [33,163]. This covert access will allow for long-term exploitation and manipulation of production processes, which can lead to quality issues, safety hazards, or intellectual property theft.

# 5 Vulnerability characterization and identification illustrative example

This section presents an illustrative example demonstrating the proposed cyber-physical defense-in-depth model and vulnerability identification approach. Section 5.1 presents an illustrative smart manufacturing system, a corresponding threat model is discussed in Section 5.2, and Section 5.3 presents the cyber-physical defense-in-depth model. Section 5.4 presents identified vulnerabilities and maps them to the different stages of the attack kill chain framework. Section 5.5 discusses how the identified vulnerabilities were exploited in empirical cyber-physical attacks. Additionally, potential vulnerability mitigation strategies are explained in Section 5.6.

## 5.1 System description

The example shown in Figure 10 represents the cyber-physical manufacturing system of a medium-sized manufacturing organization operating as a Manufacturing-as-a-Service (MaaS) provider. MaaS characterizes the increased flexibility and digital nature of the industry empowered by Industry 4.0 technologies and digital transformation, which offers on-demand availability of manufacturing capabilities and resources. Customers can upload and submit the design file (e.g., CAD or .STL files) to the manufacturer with specific GD&T requirements through a web portal, and those digital files are then stored in the cloud. A designer checks the design and product specifications, creates a computer-aided process plan, and stores the Computer Aided Process Planning (CAPP) file in the cloud. Digital files in the cloud are shared with multiple business operations, such as production facilities and post-production inspection. Operators can access and download design files and instructions regarding various production processes to controller computers on the shop floor, where parts/products are manufactured and assembled. A suite of sensors monitors production processes in real time. Following a predefined acceptance sampling plan, the post-production inspection scheme measures several key product quality characteristics, such as dimensions, locations of geometric features, and surface finish. It compares the observed results with the GD&T information stored in the Cloud. A worker monitors the system's status through an HMI and updates the machine set-up if needed. Finished products are shipped to the customer after the post-production inspection. All entities in this production system, including machine workstations, assembly stations, and post-production inspection equipment, are connected via the wireless network communication system.

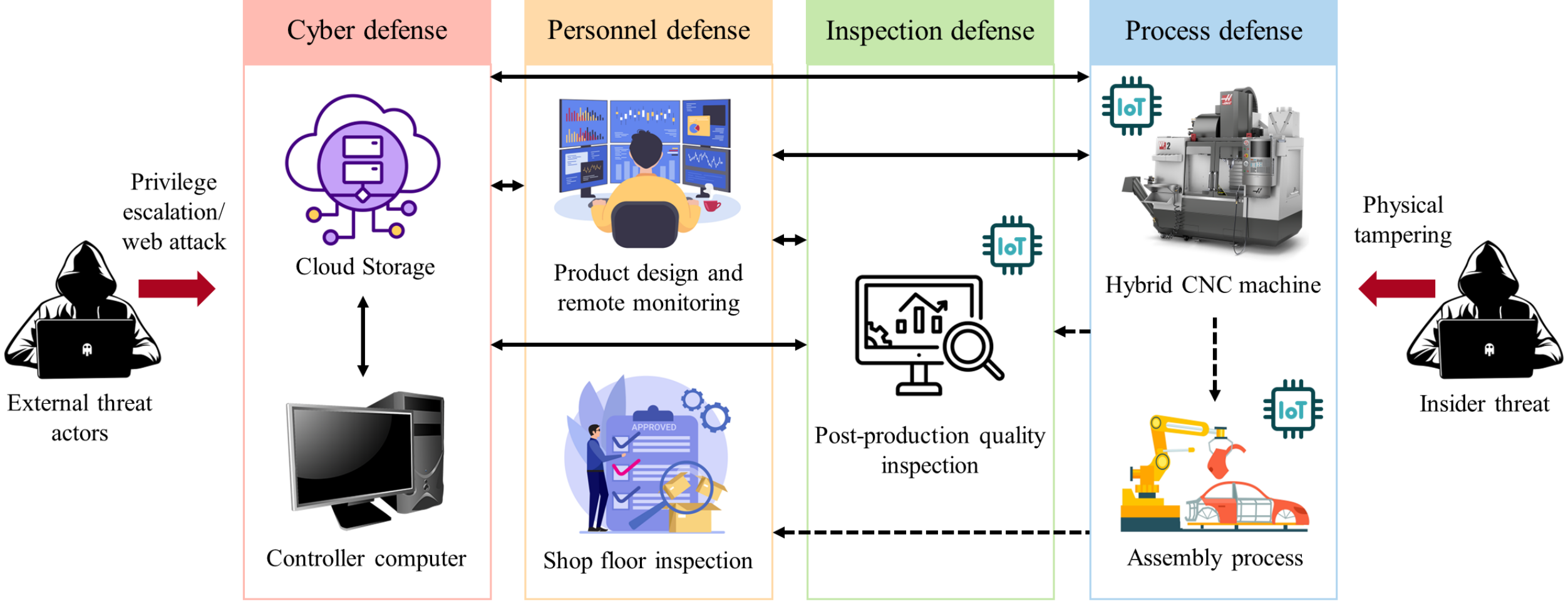

**Figure 10** An illustrative smart manufacturing system showing data flow in solid lines and material flow in dashed lines. The threat model depicts how threat actors can target different components of the manufacturing systems using various attack methods. The four blocks present different defense layers following the cyber-physical defense-in-depth model.

### 5.2 Threat model

External threat actors can target the network communication system and/or cloud storage to access and modify digital files such as CAD and tool path files to tamper with parts' dimensional, geometric, and functional integrity [54]. For example, consider the connecting rod – a critical component in piston engines – shown in Figure 11 (left), which has one hole for the rod bush bearing and two for the connecting rod bolt. A product-oriented attack can add an extra slot in the I-beam and/or remove the fillet near the hole for the rod bolt. The added slot can compromise the part's structural integrity and degrade the connecting rod's strength. Additionally, removing the fillet will create crack initiation and stress concentration and lead to potential engine failure [164]. For additively manufactured gears, inserting voids in this region will dramatically reduce the part's strength, eventually causing the failure of the part.

Additionally, a disgruntled employee can turn against a manufacturing company out of a personal grudge or for financial gain. Insiders often have unrestricted access to physical manufacturing assets and sensitive information regarding product and process specifications and the company's business operations. With such access, an insider threat can (1) steal confidential information and (2) sabotage operations targeting production processes, products, and the manufacturing ecosystem. They can exploit the lack of authentication and poor security policies to read, modify, and/or maliciously download sensitive and confidential data about products and processes. They can also tamper with production processes on the shop floor and/or disrupt operations across the manufacturing value chain.

### 5.3 Cyber-physical defense-in-depth

For the smart manufacturing system mentioned in Section 5.1, traditional cybersecurity measures such as secure data transmission, firewall, intrusion prevention system, intrusion detection system, appropriate authorization, and access control will constitute the cyber defense layer to prevent adversaries from accessing and altering the design file of a part/product while the file is in cloud storage or being transferred to the controller computer. For example, next-generation firewalls can be implemented to examine network traffic, filter out malicious websites, and prevent internet-based malware [113]. If adversaries breach the cyber defenses and gain access to the system given enough time and resources, trained personnel can add an additional defense layer against potential attacks. Training personnel on computer security should emphasize the recognition of phishing efforts, securing sensitive data, and maintaining best practices for managing passwords to prevent unauthorized entry to data [112]. Operators on the shop floor need training on the realization and identification of attack-induced alterations to products and processes, proper security tool usage, and the corresponding protocols to put in place for timely detection and quick response toward potential threats. Continuous training with simulated attack scenarios can keep employees alert and ready to protect sensitive data and operational systems in case of an attack. If the personnel cannot prevent and/or detect potential attacks, post-production inspection tools offer another avenue for attack detection. Inspection layer defenses include establishing and monitoring key security characteristics to verify product quality, introducing randomness to the design and implementation of QC tools, and developing a holistic quality signature unique to product and process designs [143]. Finally, process-layer defenses focus on attack detection and mitigation at the production process level. Appropriate detective countermeasures can be developed utilizing process dynamics, leveraging the understanding of cyberattack-induced behavioral changes in the process. For example, monitoring in-process variables such as acoustics, vibration, and power consumption can verify parts' dimensional and geometric integrity [6,135]. Organizational policies and procedures will include clear guidelines for secure operations, implementing defense measures discussed above, and incident response.

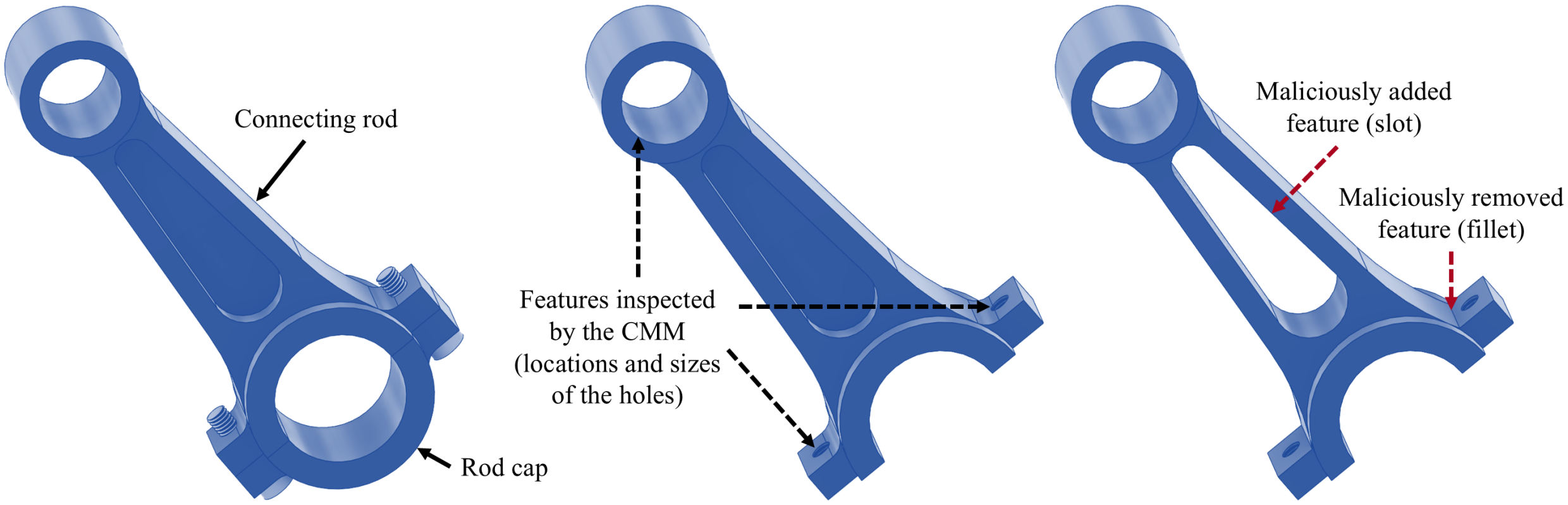

**Figure 11** Connecting rod used in piston engines (left), inspection outline for the intended product (middle), and the altered product (right)

## 5.4 Vulnerability identification and characterization

The proposed cyber-physical defense-in-depth model-driven framework enables the systematic identification of two groups of vulnerabilities. First, the MaaS provider may find that there are missing defense layers from the cyber-physical defense-in-depth model. For example, the inspection process may not be designed as a physical defense layer to detect a C2P attack. In such cases, a significant vulnerability for a manufacturing system is the absence of any defense layers discussed in the defense model. Second, each defense layer may contain vulnerabilities that adversaries can exploit. The following subsections describe identified vulnerabilities and map them to the different stages of the cyber kill chain framework.

### *5.4.1 Identified vulnerabilities*

In the cyber defense layer, an insecure web surface in the web portal – where customers upload the design file – will allow external threat actors to intercept, alter, and steal sensitive information during data transmission. Insiders can also exploit a lack of authentication and identity management to steal and tamper with digital files. Untrained employees can inadvertently download malicious files from an adversary posing as a legitimate customer and fall victim to phishing attacks. This can provide adversaries with login credentials and grant them unauthorized access to the system. Additionally, personnel may fail to diagnose the attack-induced changes to products and processes due to limited cognitive sensitivity, which causes them to fail to detect attacks. The post-production inspection system is also vulnerable due to inadequate data collection, as it only measures the key quality characteristics of a product. Even if anomalies are detected, the reason can be attributed to errors in production processes instead of cyberattacks. Insufficient physical access control to the production facility can allow threat actors to collect side-channel emissions to steal IP, reconstruct the object, and tamper with the equipment hardware and sensors, introducing vulnerabilities to the process defense layer. Additionally, compromised hardware, such as unauthorized commands/codes embedded in firmware, can allow adversaries to gain control of the equipment. Table 2 summarizes the identified vulnerabilities.

Consider the connecting rod as a sample product manufactured by the MaaS provider and the product-oriented attack described in Section 5.2. The manufacturer uses a predefined post-production inspection scheme to inspect the locations and sizes of the two holes in the connecting rod using a CMM, where the CMM uses a set of hit points to identify their locations and measure the diameters, as shown in Figure 11 (middle). Product-oriented attacks can add and delete features to and from the product to tamper with its dimensional and structural integrity, as illustrated in Figure 11 (right). However, the post-production inspection scheme based on limited CMM hit points will never notice these malicious changes as the manufacturer collects data for only a subset of features. While the design changes may seem obvious in this example, similar attacks on mass-produced products with increased complexity and design alterations with varying scales will be challenging to detect in an automated inspection environment due to incomplete or poor-quality data collection [21].

**Table 2** Identified vulnerabilities in the illustrative smart manufacturing system

| Cyber defense layer vulnerabilities | Personnel defense layer vulnerabilities | Inspection defense layer vulnerabilities | Process defense layer vulnerabilities |
|---|---|---|---|
| Insecure web interface [88] | Lack of cybersecurity awareness, knowledge, and skills [116] | Incomplete data collection and data quality [143] | Sensor tampering and interference [150] |
| Lack of authentication and identity management [91] | Limited cognitive sensitivity [15] | Data analysis deficiencies [20] | Compromised hardware [33,162] |

#### *5.4.2 Vulnerability mapping to the cyber kill chain framework*

The insecure cloud interface and/or poor authentication measures enable threat actors to perform *reconnaissance*, weaponization, and delivery stages of the cyber kill chain framework. Specifically, threat actors may covertly gather intelligence about sensitive digital assets and manipulate CAD or toolpath files to initiate the cyber-physical attack. Vulnerabilities within personnel layers, such as insufficient cybersecurity training and low cognitive sensitivity, primarily contribute to the *exploitation* and *installation* stages by making it easier for threat actors to gain unauthorized access (e.g., through phishing attacks) and persist in the system while evading detection. Incomplete data collection and analysis deficiencies can be exploited during the *exploitation* stage when attackers actively introduce changes that inspection systems fail to detect. Process defense layer vulnerabilities, i.e., compromised hardware and sensor tampering, can facilitate the *command & control* stage by enabling remote manipulation of manufacturing processes and support the *actions on objectives* stage by executing unauthorized modifications, sabotage, or data exfiltration within the system.

### 5.5 Empirical cyber-physical attacks exploiting identified vulnerabilities

This section presents several empirical cyber-physical attacks demonstrating real-world exploitation of the vulnerabilities identified in Section 5.4.

#### *5.5.1 Cyber-physical attack case studies*

Multiple academic studies have demonstrated product-oriented attacks, such as altering the part's geometric integrity and inserting voids in a component in stress concentration areas in additive manufacturing. This results in a significant loss of the part's strength and functional performance. For example, Wells et al. (2014) illustrated that digital files could be intercepted and altered during file sharing, causing premature failure of the end product [20]. Strum et al. (2017) demonstrated the introduction of internal voids by tampering with the .STL file, which significantly reduced the tensile strength of the printed part [14]. Belikovetsky et al. (2017) presented a sabotage attack on manufactured parts in which they remotely compromised the design file of a quadcopter propeller [16]. Shafae et al. (2019) demonstrated six attack scenarios that can evade traditional post-production inspection techniques and maliciously alter the geometric integrity of machined components [15]. Additionally, cyber-physical attacks on manufacturing systems can target the manufacturing equipment [11,12] and/or the integrated manufacturing ecosystem and sub-systems [17–19]. Graves et al. (2021) demonstrated an attack on manufacturing equipment where the powder delivery system was targeted for sabotaging a Powder Bed Fusion (PBF) additive manufacturing process [11].

### *5.5.2 Exploited vulnerabilities*

The empirical attacks presented in Section 5.5.1 exemplify how cyber-physical attacks can be strategically designed to exploit vulnerabilities across multiple defense layers. At the cyber layer, threat actors used insecure cloud interfaces and insufficient authentication mechanisms to covertly access and modify critical digital files like CAD models and toolpaths [14,16]. Within the personnel layer, several attacks took advantage of employees' lack of cybersecurity awareness, insufficient training, and cognitive sensitivity, limiting their ability to detect subtle but significant changes in geometric features and process parameters [15,20]. At the inspection layer, adversaries effectively exploited inherent flaws in automated quality control systems, specifically routine inspection protocols that only evaluate KQCs while ignoring changes in non-KQC features [15,21]. Table 3 briefly presents the exploited vulnerabilities and cyber kill chain steps demonstrated by these attacks. While discussing each attack in detail is beyond the scope of this study, one empirical attack scenario is presented below to illustrate how these vulnerabilities can be leveraged in practice.

**Table 3** Empirical examples of cyber-physical attacks on manufacturing systems, exploited vulnerabilities, and demonstrated steps in the cyber kill chain framework

| References | Exploited vulnerabilities | Cyber kill chain framework steps |
|---|---|---|
| Wells et al. (2014) [20] | Lack of authentication; lack of cybersecurity awareness, knowledge, and skills; limited cognitive sensitivity | Installation, command & control |
| Strum et al. (2017) [14] | Lack of authentication; lack of cybersecurity awareness, knowledge, and skills; data analysis deficiency | Installation, command & control |
| Belikovetsky et al. (2017) [16] | Compromised hardware; incomplete data collection and data quality | Reconnaissance, exploitation, command & control |
| Shafae et al. (2019) [15] | Limited cognitive sensitivity; incomplete data collection and data quality; data analysis deficiencies | Reconnaissance, exploitation, command & control |
| Graves et al. (2021) [11] | Compromised hardware; limited cognitive sensitivity; incomplete data collection | Installation, command & control |

### *5.5.3 Case study: cyber-physical attack on CNC Machining*

Shafae et al. (2019) systematically categorized and described the critical elements of potential product-oriented cyberattacks on machining systems [15]. They proposed an attack design and designation scheme and demonstrated six attacks on the geometric integrity of machined parts that can evade traditional post-production inspection techniques. A hydraulic spool valve – a critical component in various hydraulic valves used in landing gear assemblies in modern aircraft – was used as the test part for experimental investigation. In one attack scenario, the middle section of the spool valve was shifted by 2 mm to the right to degrade the functional integrity and performance of the part. Figure 12 (left) presents the nominal and altered design of the spool. This attack was aimed at a manufacturer that only inspected 25% of its spools for the correct location of the middle section. The manufacturer relied on an industrial control monitoring system utilizing machine controller data to monitor various machine and process status metrics, including average power consumption and the execution time of G-code programs.

The threat actor gained information about the product and/or post-production inspection scheme through reconnaissance before launching the attack. In doing so, adversaries can exploit cyber domain vulnerabilities such as insecure cloud interfaces and web applications and insufficient authentication and authorization to gain access to the system. Knowledge of the product and process gained through reconnaissance was exploited to minimize the detectability of the attack's physical impact on the product and process during the attack execution. First, the introduced physical change magnitude was very small compared to the original part size. Second, the attack only targeted the parts that would not be inspected according to the sampling plan deployed by the manufacturer. Third, the geometrical change in this attack did not involve changing any process-independent cutting parameters, so the average amplitude of the dependent variable (e.g., cutting power) remained constant.

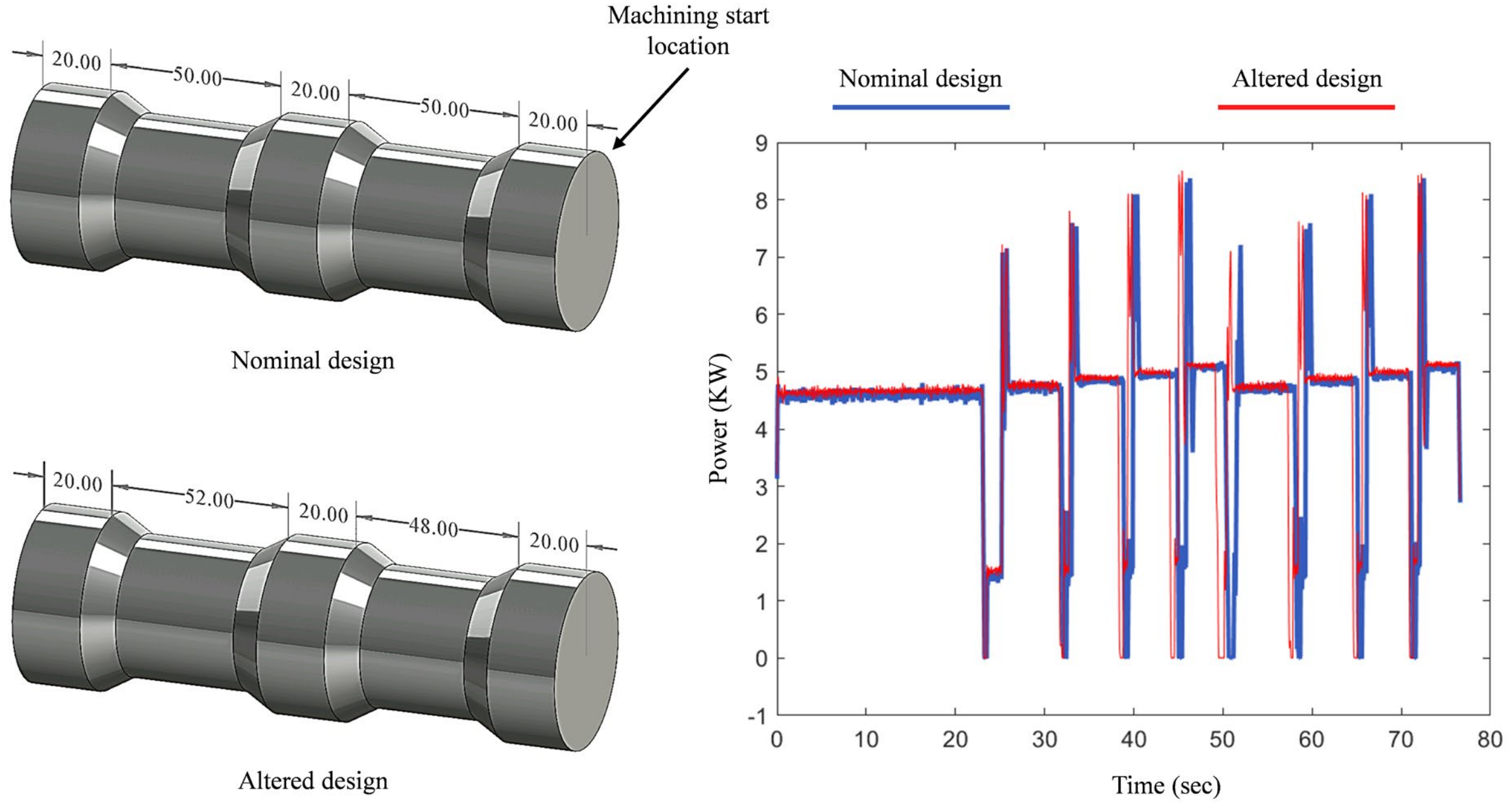

**Figure 12**: Nominal and altered geometry of a simplified spool design where all dimensions are in mm (left) and the corresponding spindle power consumption signal collected during the turning operation (right)

This attack exploits limited cognitive sensitivity, as the introduced physical change is small relative to the original part dimensions, making it difficult for shop floor operators to detect visually or through standard inspection procedures. It also leverages incomplete data collection, as the manufacturer inspects only 25% of the spools, following a predefined sampling strategy. Finally, data analysis deficiencies in process monitoring resulting from only monitoring aggregated features, such as average power consumption and total execution time, enable the attack to remain undetected. The corresponding spindle power consumptions during the rough cutting operations for the two designs using a CNC turning machine are shown in Figure 12 (right). The power profile represents nine cutting cycles: the first cutting cycle corresponds to a straight turning across the entire length of the workpiece, and the remaining eight cycles correspond to four cutting cycles per recess. As the attack does not affect process-independent cutting parameters, the steady-state spindle power (KW) and maximum spindle power (KW) remain constant for nominal and altered designs. The malicious design change subtly alters local machining times – reducing cutting time for the first recess while increasing it for the second – without affecting the overall process time. As a result, G-code execution time also remains unchanged. Therefore, the attack successfully evades detection, bypassing both post-production inspection and real-time process monitoring.

### 5.6 Potential vulnerability mitigation strategies

While a comprehensive discussion of vulnerability mitigation strategies is beyond this work's scope, several effective countermeasures are briefly described in this section for completeness. In the cyber domain, zero-trust architecture, multi-factor authentication, and network segmentation can help to reduce vulnerabilities such as insecure web interfaces and weak authentication. For example, network segmentation can prevent threat actors from accessing critical systems and data through lateral movement in the network by enforcing rule-based access controls and limiting network traffic [54]. To enhance data integrity and traceability, blockchain technology provides a tamper-proof, decentralized security framework for manufacturing data [165]. Blockchain-based version control can also ensure the integrity of digital files (e.g., CAD files), process logs, and inspection records by preventing unauthorized changes. Smart contracts can automate authentication and protect digital files [166], while watermarking techniques incorporate security features into CAD models to prevent intellectual property theft and reverse engineering [57]. Additionally, Intrusion Detection Systems (IDS) can be deployed to detect unauthorized data access, misuse, and alterations by monitoring malicious user behavior, system activity, and network traffic [167].

Cybersecurity training programs should be integrated into workforce development to address personnel-related vulnerabilities, as recent industry reports have identified security awareness as one of the most effective countermeasures against manufacturing cyberattacks [168]. Behavioral biometric authentication and AI-driven phishing detection can further reduce insider threats and social engineering risks [169]. For inspection vulnerabilities, potential mitigation strategies include (1) implementing adaptive QC strategies, i.e., introducing randomness to the design and implementation of QC tools, (2) selecting appropriate QC tools, (3) verifying the underlying statistical assumptions before implementation, (4) monitoring security-aware alternate KQCs, and (5) introducing part and product specific tags/signature during production for improved visibility and traceability [21].

Physics-informed and security-aware process monitoring approaches can be used to detect attack-induced product and process anomalies by analyzing manufacturing process dynamics variables like acoustics [170], vibration [171], and power consumption [6]. Effective methods can be developed by utilizing signal features that are sensitive and responsive to the transient anomalies induced by cyber-physical attacks instead of relying solely on features designed for traditional process monitoring. In the attack example presented in Section 5.5.3, commonly used features like the average power consumption, maximum and minimum amplitude in the power signal, and total machining time cannot distinguish between the two signals [15]. However, the difference can be observed in the local machining time for each feature, i.e., the two recesses. Therefore, extracting and analyzing local and granular temporal and amplitude signal features during process monitoring can improve attack detection likelihood [6]. Additionally, Physically Unclonable Functions (PUFs) can create unique process "fingerprints," enabling tamper detection and integrity validation [172]. For mitigating hardware-based vulnerabilities, hardware fingerprinting and tamper-resistant authentication can be designed and implemented to verify equipment integrity.

# 6 Conclusion

System vulnerabilities influence cyberattack pervasiveness and the likelihood of an attack's success, governing the organizational cybersecurity risk landscape. Hence, understanding, identifying, and categorizing vulnerabilities are essential to defend critical manufacturing infrastructure against the growing cybersecurity threat. This will allow organizations to analyze attack attributes for risk assessment, proactively address vulnerabilities, and develop and deploy effective mitigation strategies. To fulfill this need, this work systematically characterizes manufacturing-specific vulnerabilities, provides a structured classification scheme for vulnerability identification, and creates the first taxonomy for manufacturing-specific vulnerabilities.

First, it defines cyber-physical vulnerabilities in manufacturing systems and proposes the concept of vulnerability and defense duality for vulnerability characterization. Unlike current approaches primarily focusing on software and network vulnerabilities, this paper considers the unique physical and cyber-physical characteristics of smart manufacturing systems and provides a more comprehensive analysis of manufacturing-specific vulnerabilities. Second, the paper introduces the cyber-physical defense-in-depth model that depicts how potential non-cyber defense layers can be developed and deployed in addition to the traditional cyber-domain defenses. The current design and implementation of these defenses are analyzed to systematically identify vulnerabilities. Third, potential vulnerabilities in the manufacturing cyberspace, human element, post-production inspection system, and production process monitoring are identified by surveying relevant literature, industry reports, and existing vulnerability databases. This work is the first to present taxonomical classifications of manufacturing-specific vulnerabilities. Finally, the proposed cyber-physical defense-in-depth model and vulnerability identification framework are demonstrated through an illustrative smart manufacturing system and its corresponding threat model by systematically identifying vulnerabilities, mapping them to different stages of the attack chain, and analyzing how these are exploited in real-world cyber-physical attack scenarios.

Given the resource constraints small and medium enterprises (SMEs) face, integrating the developed taxonomy into existing cybersecurity audit frameworks, such as those provided by NIST [173], could enhance its practical applicability. Additionally, future work will explore the development of an audit questionnaire based on the developed vulnerability taxonomy to provide SMEs with a structured, resource-efficient approach to assessing cyber-physical security risks. The presented cyber-physical defense-in-depth model and reported vulnerabilities will pave the way toward developing security-aware personnel, security-aware inspection, and security-aware processes, laying the foundation for more secure cyber-physical manufacturing systems. This work establishes a foundation for the systematic identification and classification of cyber-physical vulnerabilities in manufacturing systems, offering researchers and practitioners a comprehensive understanding of how adversaries can exploit these weaknesses. Future research should build on this foundation by investigating how such vulnerabilities facilitate different stages of the cyber kill chain, thereby supporting informed threat modeling and the development of targeted mitigation strategies.

## Acknowledgment

This work has been partially funded through the Technology and Research Initiative Fund (TRIF) under the National Security Systems Initiative, funded under Proposition 301, the Arizona Sales Tax for Education Act.